\documentclass[fleqn,usenatbib]{mnras}

\usepackage{newtxtext,newtxmath}
\usepackage[dvipsnames]{xcolor}
\usepackage[T1]{fontenc}
\DeclareRobustCommand{\VAN}[3]{#2}
\let\VANthebibliography\thebibliography
\def\thebibliography{\DeclareRobustCommand{\VAN}[3]{##3}\VANthebibliography}
\usepackage{rotating} 
\usepackage{graphicx}	%
\usepackage{amsmath}	%
\usepackage{booktabs}
\usepackage[normalem]{ulem} %
\usepackage{subcaption}
\usepackage{float}
\usepackage{grffile}
\usepackage[version=4]{mhchem}
\usepackage{adjustbox}

\usepackage{array}
\newcolumntype{x}[1]{>{\centering\let\newline\\\arraybackslash\hspace{0pt}}p{#1}}
\graphicspath{{figures/}}

\def \kms  {$\rm km\,s^{-1}$}

\def \mum {$\mu$m\,}

\def\purple#1 {{\textcolor{purple}{#1}}\ }
\def\red#1 {\textcolor{red}{#1}}
\def\new#1 {{\bf #1 }}
\def\blue#1 {{\textcolor{blue}{#1}}\ }
\newcommand\orcid[1]{\href{http://orcid.org/#1}{\adjustbox{trim={-.15\width} {0\height} {-.15\width} {0\height},clip}{\includegraphics[height=12pt]{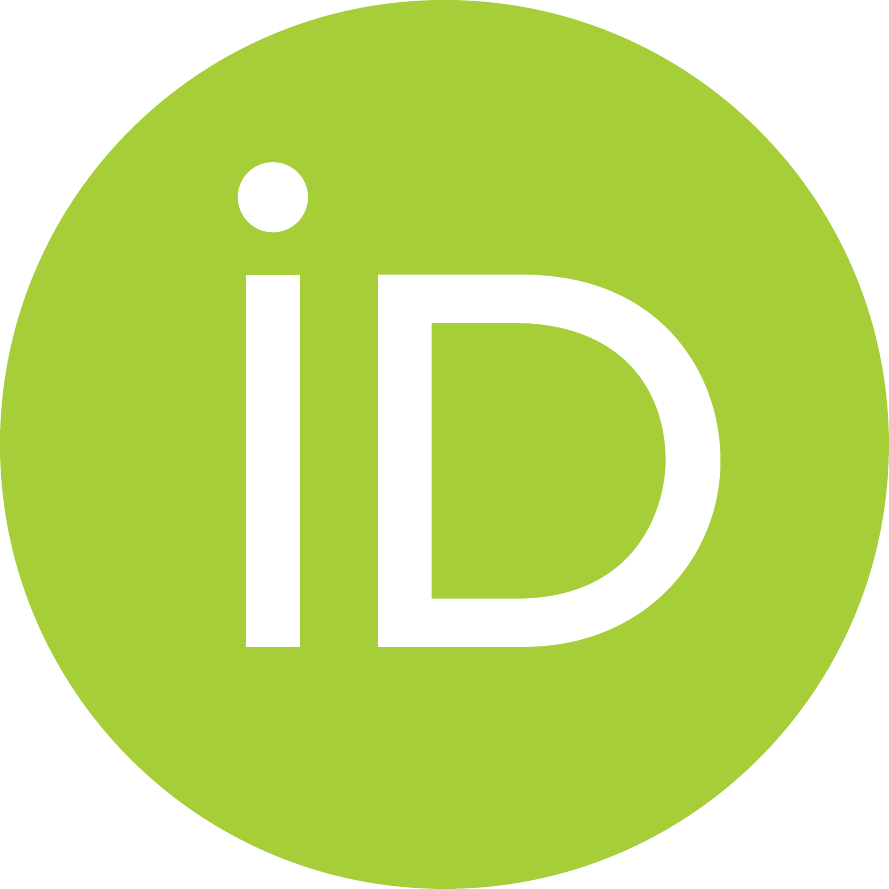}}}}

\title[VLT/KMOS Observations of IC~443]
{Multiple gas phases in supernova remnant IC 443: mapping shocked \ce{H2} with VLT/KMOS}

\author[Y. Deng et al.]{
\Large
Yunwei Deng\orcid{0000-0002-7478-6427}\(^{\! 1,2\thanks{E-mail: dengyw@smail.nju.edu.cn}}\),
Zhi-Yu Zhang\orcid{0000-0002-7299-2876}\(^{\! 1,2\thanks{E-mail: zzhang@nju.edu.cn}}\),
Ping Zhou\orcid{0000-0002-5683-822X}                              \(^{\! 1,2}\),    
Junzhi Wang\orcid{0000-0001-6106-1171}                            \(^{\! 3}\),  
Min Fang                               \(^{\! 4}\), 
\newauthor
Lingrui Lin\orcid{0000-0002-2231-8381}                            \(^{\! 1,2}\),  
Fuyan Bian                             \(^{\! 5}\),    
Zhiwei Chen                            \(^{\! 4}\),   
Yong Shi\orcid{0000-0002-8614-6275}  \(^{\! 1,2}\),   
Guoyin Chen                            \(^{\! 1,2}\),  and
Hui Li\orcid{0000-0002-1253-2763}       \(^{\! 6\thanks{NASA Hubble Fellow}}\)
\\
\(^1\)  School of Astronomy and Space Science, Nanjing University, Nanjing 210093, People’s Republic of China\\
\(^2\)  Key Laboratory of Modern Astronomy and Astrophysics (Nanjing University), Ministry of Education, Nanjing 210093, People’s Republic of China \\
\(^3\)  Guangxi University, 179 Mingxiudong Road, Nanning City, Guangxi 530004, China \\
\(^4\)  Purple Mountain Observatory, Chinese Academy of Sciences, 10 Yuanhua Road, Nanjing 210023, China\\
\(^5\)  ESO,Vitacura Alonso de C\'{o}rdova 3107 Vitacura, Casilla 19001 Santiago de Chile, Chile \\
\(^6\)  Department of Astronomy, Columbia University, New York, NY 10027, USA
}

\date{Accepted 2022 October 28. Received 2022 October 28; in original form 2022 June 30}
\usepackage{multirow}
\pubyear{2022}

\begin{document}
\label{firstpage}
\pagerange{\pageref{firstpage}--\pageref{lastpage}}
\maketitle

\begin{abstract}
Supernovae and their remnants provide energetic feedback to the ambient
interstellar medium (ISM), which is often distributed in multiple gas phases.
Among them, warm molecular hydrogen (\ce{H2}) often dominates the cooling of the
shocked molecular ISM, which has been observed with the \ce{H2} emission lines at near-infrared
wavelengths. Such studies, however, were either limited in narrow filter
imaging or sparsely sampled mid-infrared spectroscopic observations with
relatively poor angular resolutions. Here we present near-infrared ($H$- and
$K$-band) spectroscopic mosaic observations towards the A, B, C, and G regions
of the supernova remnant (SNR) IC 443, with the K-band Multi-Object Spectrograph
(KMOS) onboard the Very Large Telescope (VLT). We detected 20 ro-vibrational
transitions of \ce{H2}, one H line (Br$\gamma$), and two [Fe {\sc ii}]
lines, which dominate broadband images at both $H$- and $K$-band. The
spatial distribution of \ce{H2} lines at all regions are clumpy on scales from
$\sim 0.1$ pc down to  $\sim 0.008$ pc. The fitted excitation temperature of
\ce{H2} is between 1500 K and 2500 K, indicating warm shocked gas in these
regions. The multi-gas-phase comparison shows stratified shock structures in all
regions, which explains the co-existence of multiple types of shocks in the
same regions. Last, we verify the candidates of young stellar objects
previously identified in these regions with our spectroscopic data, and find
none of them are associated with young stars.  This sets challenges to the
previously proposed scenario of triggered star formation by SNR shocks in
IC~443.

\end{abstract}

% We compare with neutral gas and cold \ce{H2} traced by H {\sc i} and \ce{CO},
% with synergy of different species in multi-gas phases. 
% 
% 
% These help us to reveal how gas in the molecular clumps
% gets excited and emits in interaction with shocks and derive partition of the
% shocked warm \ce{H2}. 
% 


%%%%%%%%%%%%%%%%%%%%%%%%%%%%%%%%%%%%%%%%%%%%%%%%%%

%%%%%%%%%%%%%%%%% BODY OF PAPER %%%%%%%%%%%%%%%%%%
\section{Introduction}

Supernova feedback to the interstellar medium (ISM) plays a key role during the
evolution of stars and galaxies
\citep{2014MNRAS.445..581H,2020ApJ...905...35K}. Massive stars are born from
molecular clouds and interact with their ambient natal ISM throughout their
lives \citep{2003ARA&A..41...57L,2019ARA&A..57..227K}.  Core Collapse Supernova
remnants (SNRs), as the end of massive stars, play a significant role in
these processes, generating large amounts of shock, radiation, and
heavy metals. These processes would compress, heat up, dissociate, and ionize
their surrounding molecular clouds (MCs), and distribute heavy elements
throughout the galaxies.  SNRs are also important sources of accelerating
cosmic rays which can heat the dense cores of the MCs. Interaction between SNRs
and giant molecular clouds (GMCs) may enhance
\citep{1984MNRAS.207..909I,2013MNRAS.429..189L} or suppress
\citep{2019MNRAS.488.4753D} sequential star-formation, and then regulate the
evolution of galaxies.

Supernova shock provides energetic feedback to the ISM
\citep[][]{1993ARA&A..31..373D}. These shocks carry a significant amount of
mechanical energy which gradually dissipate in turbulence as the shocks
propagate through and interact with the ambient ISM. They can heat, compress
and accelerate the gas in GMCs and may turn the gas into different phases
(\ce{H2}, H {\sc i}, and H$^+$) with a wide range of shock parameters: of
several $10-10^4 \text{ km s}^{-1}$ in velocity, $10-10^5\text{ cm}^{-3}$ in
volume density, and $10-10^8$ K in temperature \citep[][]{1980ARA&A..18..219M}.
These processes lead to a wide variety of structures and observables.

IC~443, also named as the Jellyfish Nebula from its optical morphology, is a
mixed-morphology SNR located in the galactic anti-centre \citep[see][for
references]{1998ApJ...503L.167R,Jones1998}. IC~443 has a diameter of $\sim50'$
at the optical and radio bands and is thought to contact with the Gem OB1
association at a distance of $\sim$1.5-2.0 kpc \citep{1995ApJ...445..246C}. We
adopt 1.6 kpc from the measurement of {\it Gaia} Data Release 2 (DR2) parallax and dust extinction
\citep{2020A&A...633A..51Z}. IC~443 originated from a core-collapse supernova
explosion, with a large uncertainty in its age ($\sim 3$--$30 \times10^3$\,yr)
\citep{1999ApJ...511..798C, 2001ApJ...554L.205O, 2008A&A...485..777T}, 

%Located in the galactic anti-centre
%direction, there are few other molecular clouds in its line of sight
%\citep{2001ApJ...547..792D}. 

%in the northwestern and southeastern, 

With clear evidence of impacts with nearby surrounding MCs, IC~443 provides a
unique test field of shock conditions in the Milky Way. The shock processes are
not only accelerating and stratifying molecular gas, but also powering atomic
and ionized gas phases with strong and extremely broad emission lines. These
shock features have been detected in CO, \ce{H2O}, H {\sc i}, etc.
\citep[e.g.][]{1979ApJ...232L.165D,1992ApJ...400..203D,2005ApJ...620..758S, 2008AJ....135..796L, 2010SCPMA..53.1357Z},
which show broadening features in almost all gas clumps, especially the B, C,
and G regions \citep[following the nomenclature of][]{1986ApJ...302L..63H}.
\ce{H2O} observations \citep{2005ApJ...620..758S} further show that a variety
of shock types and velocity components are required.

As the most abundant molecule in the ISM, molecular hydrogen is therefore the
major molecule in the shock-MC interacting regions. As a homo nuclear molecule,
\ce{H2} has no permanent dipole electric moment. Because of this, \ce{H2} can
only emit with its electric quadrupole transitions at infrared wavelengths,
that are more easily reachable with space-based instruments. Using the
long-slit mid-IR echelle spectrometer onboard NASA Infrared Telescope Facility (IRTF), \cite{1995ApJ...449L..83R}
detected \ce{H2} 0-0 S(2) rotational transition at 12.28 \mum\, towards the C
clump of IC~443, and confirmed the presence of a partially dissociating
J-shock. Based on a time-dependent shock model, \cite{1999A&A...348..945C}
measured intensities of \ce{H2} pure rotational lines with the {\it Infrared Space
Observatory (ISO)} and found a fast shock with short timescales of 1000--2000
years; \cite{2011ApJ...732..124S} proposed a scheme of the co-existence of both
C- and J-shocks that ``the C-type shocks propagate into `clumps', while J-type
propagate into `clouds' (inter-clump medium)''.

%\red{Here we should not `list' what have been done. It is always better to tell a story about the progresses in the field. I know that there are perhaps too many things to be said, and it is really difficult to make the decision. Simply get the most central logic chain that is related to our work. I will try to work it around.} 

Besides the pure rotational transitions, the rotational-vibrational (hereafter
{\it ro-vibrational}) transitions of \ce{H2} molecules are also often seen in
shocked regions.  The first energy level for ro-vibrational transitions is 1-0
S(0) (${\rm v}=1$-$0$, ${\rm J}=2$-$0$) with an upper level energy of $6471.4$\,K, allowing
it to trace warm molecular gas at temperature $T_{\rm kin} \ge 10^3$\,K.

%which means vibrational transitions at non-ground rotational levels, would play a major role in the gas cooling. 

Though the warm and hot \ce{H2} can only make up a small mass fraction of all
the shocked \ce{H2}, it plays an important role as the major coolant in
molecular shocks \citep{2010MNRAS.406.1745F,2020A&A...643A.101L}. The cooling
processes become much more efficient at a higher temperature because the cooling
efficiency of \ce{H2} at $10^3$\,K is about three orders of magnitude larger
than that at $10^2$\,K \citep{1999MNRAS.305..802L}.

The shocked molecular gas of IC~443 mainly distributes in an incomplete ring
shape, with a bright $\omega$-shaped ridge in the south. It consists of
several bright clumps, namely A, B, C, D, etc., which seem to be connected with
more extended and fainter emission. \cite{2012ApJ...749...34L} interpreted
these small gas clumps as dense cores from their parental MCs, where the
progenitor star of IC~443 was formed. From narrow-band mapping observations of
\ce{H2} 1-0 S(1), \cite{1988MNRAS.231..617B} found that the \ce{H2} emission
comes from a sinuous ridge where the molecular hydrogen is directly shocked by
the supernova shock wave.  The spatial distribution of shocked \ce{H2} is
remarkably similar to the high-velocity \ce{CO}, \ce{HCO+}, and \ce{HCN}
molecules.

\cite{2019ApJ...884...81R} made
high-resolution (R=67000) observations of pure rotational lines and found
that, to best fit the observed \ce{H2} S(5) line profile, a multi-component
shocks (CJ-type shocks combined with J-type shocks at gas
densities of $10^3$--$10^4 \text{ cm}^{-3}$) are needed.

%  Starting from \cite{1995ApJ...449L..83R}, several spectroscopic observations
%  have been performed for the \ce{H2} pure rotational lines towards the southern
%  ridge
%  \citep[][]{1999A&A...348..945C,2007ApJ...664..890N,2011ApJ...732..124S,2019ApJ...884...81R}.
%  These studies show that at least two types of shocks are required to explain
%  the \ce{H2} excitation: a fast J-type shock into inter-clump gas and a slow C-type shock into clumps [citation]. 

\cite{1988MNRAS.231..617B}, \cite{2001ApJ...547..885R}, and
\cite{2020ApJ...899...49K} have mapped the \ce{H2} 1-0 S(1) and 2-1 S(1)
transitions of IC~443 with broadband or narrowband filters. Although these
imaging observations detected a significant amount of \ce{H2} ro-vibrational
lines emission, they could not distinguish the contribution from each line and
could not derive the population of \ce{H2} vibrational levels.  On the other
hand, \cite{1995ApJ...454..277R} and \cite{2011ApJ...732..124S} observed
several ro-vibrational lines in IC~443 with a long-slit spectrograph.  However,
the sizes of the slits only gave limited fields of view (FoVs) to resolve
detailed spatial structures. Integral field spectroscopic observations, which
offer both moderate spectral resolution and wide FoV into account, are required
to simultaneously map multiple \ce{H2} transitions.

Furthermore, it has been debated whether SNRs and their progenitors could
trigger star formation in their surrounding clouds on fairly short timescales.
\cite{2011ApJ...727...81X} proposed star formation would be triggered by the
expansion of IC~443, by identifying candidates of protostellar objects and
young stellar objects (YSOs), based on {\it Infrared Astronomical Satellite IRAS} Point Source Catalog and the Two
Micron All Sky Survey \citep[2MASS,][]{2006AJ....131.1163S}. Similarly,
\cite{2014ApJ...788..122S} selected YSO candidates based on 2MASS and
{\it Wide-field Infrared Survey Explorer} \citep[{\it WISE},][]{2010AJ....140.1868W}
databases using a colour-colour diagram. More recently,
\cite{2020A&A...644A..64D} combined {\it Gaia}, {\it WISE}, and 2MASS databases and selected
protostar candidates with colour-colour filtering method in the IC 443 G region,
where they propose to have a higher concentration of protostar candidates in
the shocked clump.  However, all these photometric selected  YSO candidates
still need to be verified with spectroscopic observations at near-IR
wavelengths. The K-band Multi-Object Spectrograph (KMOS) onboard the Very Large
Telescope (VLT) offers an opportunity to testify if the candidates are real.

In this paper, we report near-infrared mapping results of ro-vibrational
transitions of \ce{H2}, Br$\gamma$ line, and [Fe {\sc ii}] lines over regions
A, B, C, and G in IC 443, using KMOS onboard VLT. In Section~\ref{sec:data}, we
describe information on the KMOS observations and data. We then describe the
results and analysis of the lines observed in these four regions in
Section~\ref{sec:results}. In Section~\ref{sec:discussion}, we compare our data
with mid-infrared pure rotational lines and with the distributions of neutral
gas and cold \ce{H2} traced by H {\sc i} and CO. We also discuss the
contribution of the \ce{H2} ro-vibrational lines at $K_{\rm s}$ band and falsify the
YSO candidates in the G region. Finally, in Section~\ref{sec:conclusion}, we
make a brief conclusion and prospect our future design about our
multi-wavelength campaign which is aiming to build a panchromatic perspective
of IC 443 from the observational aspect by combining our \ce{CO}, H {\sc i},
and H$\alpha$ data.

\begin{figure*}
\includegraphics[width=2\columnwidth]{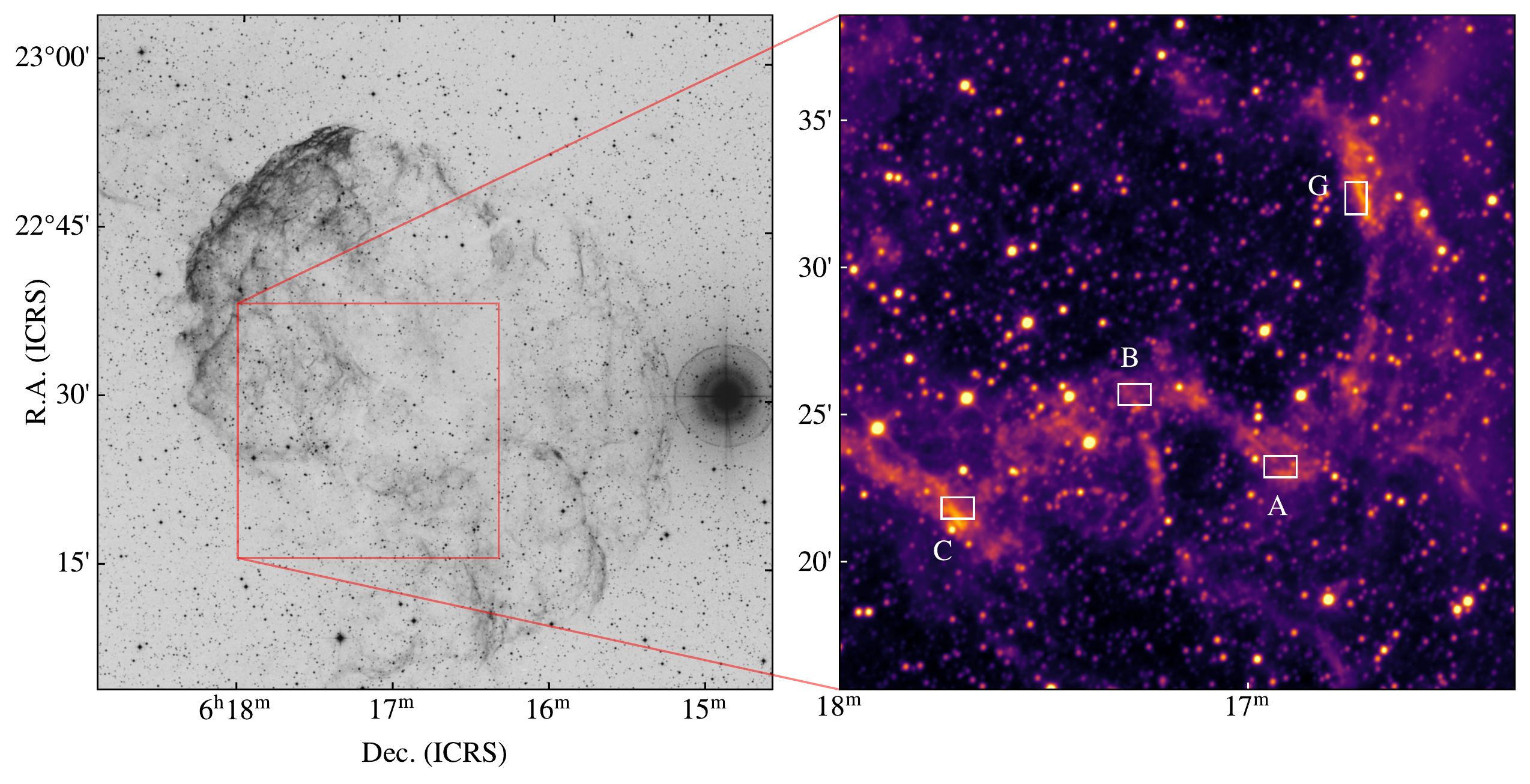}
\caption{\textit{left}: DSS R band image of IC~443; \textit{right}:
WISE W1 (3.6$\mu$m) image of IC~443. Four regions (A, B, C, and G) in the
southeastern molecular ridge we observed with VLT/KMOS are shown in the zoom-in
region with white boxes surrounding them.} \label{fig:fov}
\end{figure*}

\section{Observations and data reduction}
\label{sec:data}

\subsection{Choice of regions}
\label{sec:regions}

% \begin{itemize}
%   \item Purpose of the mapping observation. -- studying \ce{H2} maps in the GMC-SNR shocked regions. 
%   \item The morphology of shocked \ce{H2} gas (traced by high-velo CO lines) of IC 443. 
%   \item Nomination 
% \end{itemize}

% IC 443 has a large extend $\omega$-shaped molecular gas ridge located at its
% southern part, where shock waves are intensively interacting with molecular gas in the GMCs. 

% In the southern molecular shock region of IC~443, several molecular clumps with
% angular size about 1$'$ show broadened (up to 50 \kms) CO line emission of
% molecules along the interacting interface.
% \citep[e.g.][]{1992ApJ...386..158W,1993A&A...279..541V,2010SCPMA..53.1357Z}.
% \cite{1988MNRAS.231..617B} present that infrared \ce{H2} emission also shows
% similar clumpiness distributed along the $\omega$-shaped ridge. Infrared
% integral-field spectroscopic observations can obtain the emission spectra and
% map the spatial distribution of molecular hydrogen at the same time, thus
% facilitates us reveal supernova shock structures traced by warm molecular gas
% and resolve the clumpy regions created by the SNR-GMC interaction.
% 

We select four molecular clumps of IC~443, regions A, B, C, and G (see
Figure~\ref{fig:fov}), as our targets of KMOS observations. Among them, IC~443
B, C, and G have been detected with the most prominent broadened CO features of
angular sizes $\lesssim 1'$
\citep[e.g.][]{1992ApJ...386..158W,1993A&A...279..541V,2010SCPMA..53.1357Z}.
Although IC~443~A does not show a broad \ce{CO} emission, it has a prominent
high-velocity \ce{HCO+} emission \citep{1992ApJ...400..203D} and H$\alpha$
emission (see Section~\ref{sec:Compare}). 

From the currently known geometry and kinematic structures
\citep[e.g.][]{1993A&A...279..541V}, shocks in region A, B, and C are propagating
mostly along the line of sight and are reaching gas clumps in succession.
Shocks in the G region, on the other hand, propagate perpendicular to the line
of sight.

%propose new VLT CRIRES+ and ALMA CO 2-1 observation (beamsize $\sim$ 2-3") to better resolve the molecular clumps with higher  velocity resolution.
\subsection{KMOS data}

\subsubsection{Observations} \label{sec:Obs} 

% \begin{itemize}
%   \item Overview of the instrument and the observing scheme.   
%   \item Observing dates, modes, weather conditions, etc. 
%   \item Detailed hardware configuration and setups (resolution, integration time, etc). 
% \end{itemize}

Our observations (Program ID: 0104.C-0924, PI: Zhi-Yu Zhang) were performed
with the near-infrared K-band Multi-Object Spectrograph \citep[KMOS,][]{sharples2013first}
onboard the Very Large Telescope (VLT) at Cerro Paranal in Chile, during
December 2019 and January 2020.

KMOS has 24 configurable arms, of each an integral field unit (IFU) is
employed. Each IFU has 14$\times$14 pixels with a spatial sampling
0.2$''\times0.2''$, thus a spatial coverage of of 2.8$''\times2.8''$.  We
adopted the {\it Mosaic Mode} to carry out the mapping observations, which
arrange the 24 arms in a 6$\times$4 grid of fixed dimensions. For each scan,
the telescope move across 16 successive telescope pointings,  which allows a
complete mosaic of data covering a region of $64.9"\times43.3"$. 

We use the {\it HK} grating to cover both $H$ and $K$ bands simultaneously (from
1.46 \mum\, to 2.41 \mum). It offers a spectral resolution of $\sim 2000$,
which correspond  to a velocity resolution of $\sim165$\,km\,s$^{-1}$ at $\sim$
2 \mum.

% All nights have conditions of clear skies (or better) and thin sky
% transparency. The precipitable water vapour (PWV) were less than 20 mm and the
% seeing ranges from $0.53"$ to $2.29"$ for different mosaic exposures, \ywr{with
% a mean of 0.99$''$, one of the 16 region A mosaic exposures are taken with
% seeing larger than 2$''$}[ with a median of $\sim$ xxx?$''$ If the seeing has big
% jumps, we may perhaps make a table to show it. If only one or two OB
% had poor seeing, we can simply state it.]. 

The observations were performed with eight observing blocks (OBs).  Each mosaic
scan consists of 16 successive telescope pointings and exposures.  For each of
the 16 pointings, we spent 150\,s on source and 150\,s for sky exposure. Such
an observational configuration maximises the mapping area but all pixels have
only a single exposure, except for the pixels in the overlap regions of two
IFUs. During our observations, unfortunately, one of the 24 arms was not
functional
\footnote{\url{http://www.eso.org/sci/facilities/paranal/instruments/kmos/news.html}},
leaving a rectangular empty area in the datacube at each region (white squares in Figure~\ref{fig:H2Emission}). 

All nights have conditions of clear skies (or better) and good sky
transparency. The precipitable water vapour (PWV) was 3.0 mm on average and the
seeing ranges from $0.53''$ to $2.29''$ for different mosaic exposures, with a
mean of $0.99''$. Only one mosaic exposure in region A was taken with seeing
larger than $2''$.

\subsubsection{Data Reduction}

All mosaic data were initially reduced using the ESO KMOS pipeline
\citep[ver. \textsc{kmos-3.0.1} ][]{2013A&A...558A..56D} integrated into the
interactive pipeline operating environment ESO Reflex: ESO Recipe Flexible
Execution Workbench \citep[\textsc{esoreflex-2.11.0};][]{2013A&A...559A..96F}.
ESO Reflex can automatically perform most of the main reduction steps including
dark, flat, wavelength calibration, luminosity correction, telluric and
atmospheric correction, flux calibration, and cube reconstruction and
combination, etc. The calibration data we adopted was delivered with the
pipeline and the ESO \verb"CalSelector"
service\footnote{\url{https://archive.eso.org/cms/application_support/calselectorInfo.html}}.
We use a spatial sampling rate of $0.2''$ per pixel and a spectral sampling
rate of $0.46$\,nm per channel to reconstruct the 3-D data cubes, satisfying a
Nyquist sampling. The consequent velocity sampling rate is 84\,km\,s$^{-1}$
for 1.6440 $\mu$m [Fe {\sc ii}] line and 70 \,km\,s$^{-1}$ for 2.1218 $\mu$m
\ce{H2} 1-0 S(1) line.

% The calibrated 3-D data cubes were generated directly with physical flux and
% coordinate system on each channel and pixel. 
% However, the error for the mosaic mode was not properly propagated in the pipeline and the correct noise cube cannot be
% generated. ESO Operation Helpdesk suggests us modeling the emission or
% absorption line region and using the standard deviation of the residuals as a
% measure of noise, but the shock broadened non-Gaussian profiles of emissions
% lines leave the large uncertainty in the modeling. 

Because each pixel was observed only once, the error was not properly
propagated in the pipeline. To obtain a channel-based noise, the uncertainties
of line intensities are estimated by the standard deviation $\sigma_\lambda$ of
a moving-box that contains 20 line-free channels at both sides of each line,
and this $\sigma_\lambda$ is regarded as the uncertainties for all the line
channels. Our estimated $\sigma_{F_{\lambda}}$ is on the order of
$10^{-19}$\,erg\,s$^{-1}$\,cm$^{-2}$\,\AA$^{-1}$ per pixel per channel.  After
examining with stars observed in our data with the 2MASS catalog, we adopt an
uncertainty of 20\% in the absolute flux calibration, which is consistent with
the 5-20\% uncertainty suggested by the KMOS user manual.

%Therefore we assume that the noise is uniform and adopt the standard deviation measured between spectral channels as our noise. To obtain the channel-based noise, we first remove all signals above 3 $\sigma$ and drive the standard deviation from all rest channels that are assumed to be line-free channels.  

% The estimation of flux
% calibration uncertainty is a long-standing issue for KMOS, the KMOS user manual
% suggests an uncertainty of 5-20\% in the absolute flux calibration.  Since the
% accuracy of absolute flux does not significantly influence the key
% conclusions in this paper, all the uncertainties in this paper don't include
% the flux calibration uncertainty.

During combining all cubes produced from subscans, the mosaic tessellation
sometimes is not always perfect because of the errors of pointing, and sometimes
the pipeline makes the seams of the mosaic sub-cubes with unnaturally bright structures,
especially in regions A and G (see Figure~\ref{fig:H2Emission}). 

Therefore, we masked the apparent abruption and misplacement visually found in
the final mosaic cube. The OH line subtraction is not optimal in some cases,
especially for region G. We further subtract the median value of
nearby line-free pixels to remove the remaining radiance features. 

%\zy{Let's double check it.}

The \ce{H2O} vapour leads to an enhanced atmospheric absorption of almost zero
transmission nearby 1.87$\mu$m, depending on the PWV (See
Figure~\ref{fig:spectra}). In the following analysis, we simply abandon all lines that are severely
affected by absorption. Although some lines in this wavelength range seem to be well detected, their fluxes have too high uncertainties to be trusted.

% \subsection{JCMT \ce{CO} 3-2 data}
% \subsection{JVLA H {\sc i} data}
% \subsection{ZTF data}

\subsection{Ancillary Data}
\label{sec:ancillary}

In this work, we compare the KMOS data with H~{\sc i} 21-cm data from the Karl
G. Jansky Very Large Array (JVLA), the old VLA, and the Arecibo telescopes, CO
$J$=3-2 from the James Clerk Maxwell Telescope (JCMT), H$\alpha$ data from the
SITELLE onboard the Canada-France-Hawaii Telescope (CFHT), and \textit{r} band
data from the Zwicky Transient Facility (ZTF).

%\zy{please cite papers of these telescopes} 

%in the archive to compare the distribution
%of different phases of hydrogen in these shocked regions
%(Section~\ref{sec:Compare}, \ref{sec:Stratify}).

The VLA/JVLA 21-cm H{\sc i} data consists of multiple observations taken with
the B, C, and D array configurations. Among them, the JVLA B- and C-array
observations are performed in 2016 (Project ID: 16A-266, PI: Ping Zhou). The
D-array observation is performed in 2001 \citep[Project ID:
AK0537,][]{2008AJ....135..796L}. The Arecibo 21-cm H{\sc i} data is acquired
from the GALFA-H {\sc i} survey DR 1 \citep{2011ApJS..194...20P}. We combined
all VLA/JVLA and Arecibo data with the Total Power Map To Visibilities packages
\citep[{\sc TP2VIS},][]{2019PASP..131e4505K}, and image it with Common
Astronomy Software Applications version 6.1.2.7 (CASA)
\citep{2007ASPC..376..127M}.  Details about the observations and data reduction
will be reported in a separate paper (Lin et al. in prep.).

The CO $J$=3-2 data was observed in November 2015 with submillimetre spectral
imaging system HARP/ACSIS \cite{2009MNRAS.399.1026B} onboard JCMT (Proposal ID:
M15BI126) and we download the reduced data from the JCMT science archive. 

The CFHT/SITELLE data are performed in 2021 (Project ID: 21BS020, PI: Yunwei
Deng). SITELLE is an imaging Fourier transform spectrometer which offers
$11'\times11'$ large FoV and high-spectral resolution simultaneously
\citep{2019MNRAS.485.3930D}. IC~443 A and B were observed in the SN3 band with a
spectral resolution of $R = 9000$. We extract the moment-0 maps of H$\alpha$
emission to multi-band comparison. For the C and G regions of IC~443, we
download the ZTF \textit{r} band data from the IRSA ZTF archive
\citep{2019PASP..131a8003M} as a substitution to trace H$\alpha$ emission.

In Section~\ref{sec:YSO}, we use catalogs from Gaia eDR3
\citep{2021A&A...649A...1G} and LAMOST DR7 \citep{2015RAA....15.1095L} to
help analyze the YSO candidates.

%\footnote{\url{https://archive.nrao.edu/archive/advquery.jsp}}
%\footnote{\url{https://www.eaobservatory.org/jcmt/science/archive/}}. 
%\footnote{\url{https://irsa.ipac.caltech.edu/Missions/ztf.html}}.

%2MASS \citep{2006AJ....131.1163S} and WISE \citep{2010AJ....140.1868W} point source catalogs to

\section{Results}
\label{sec:results}

\subsection{Near Infrared emission lines in shocked regions}

\begin{figure*}

\includegraphics[width=2.15\columnwidth]{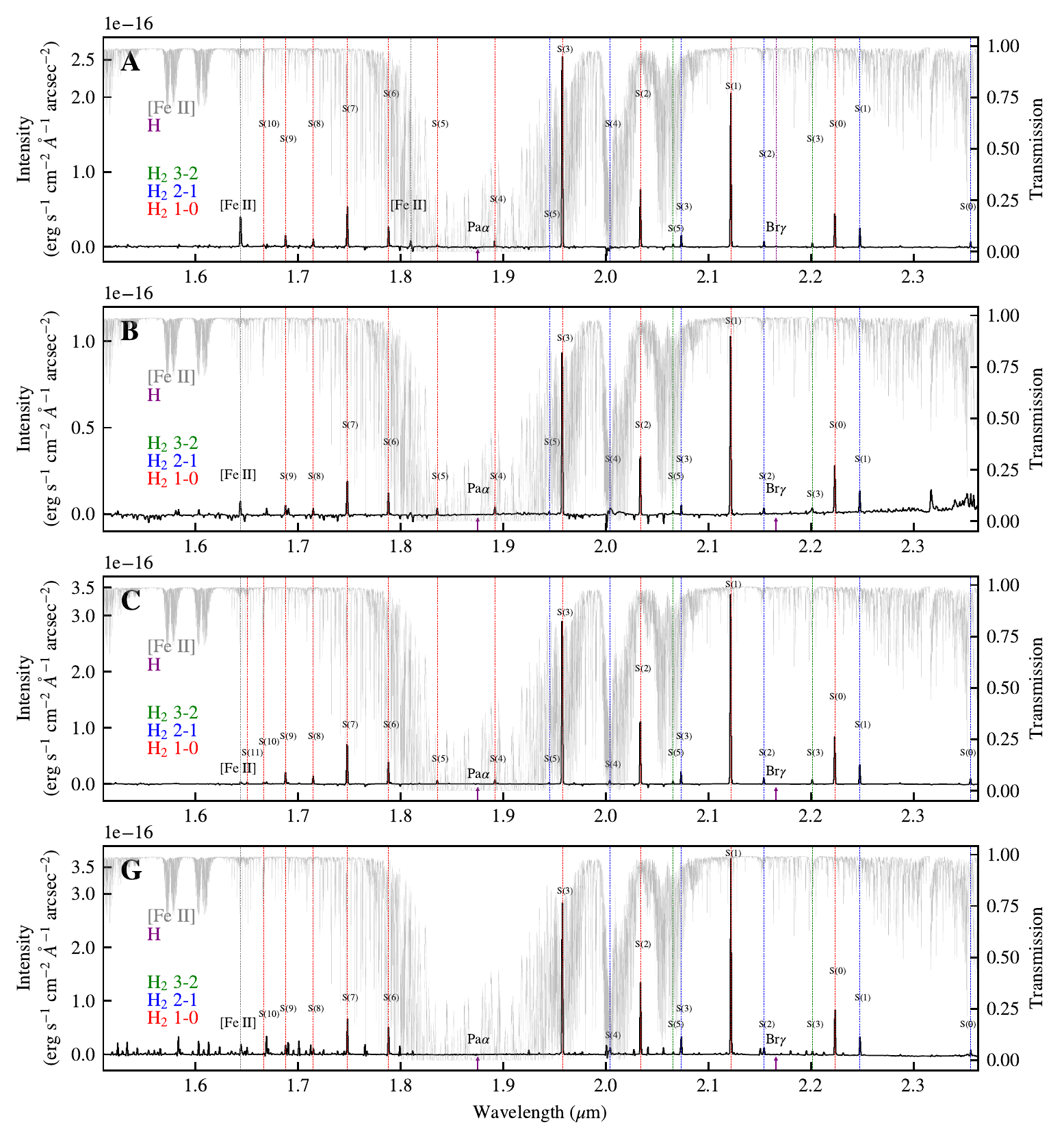}
\caption{Spatially-averaged spectra from regions of IC~443 A, B, C, and G. 
        Red, blue, and green dashed lines label \ce{H2} lines with 1-0, 2-1 and
        3-2 vibrational transitions, respectively. Grey dashed lines present
        [Fe {\sc ii}] lines, and purple arrows show the positions of the undetected  H
        atomic lines. The background grey lines are the atmospheric
        transmission at Paranal with PWV=3.5 mm obtained from ESO {\sc
        SkyCalc}. The line features without labeled out is mostly OH residuals
        from the atmosphere
        \citep[refer to Figure~\ref{fig:radiance}, and][for the OH line catalog]{2000A&A...354.1134R}. } \label{fig:spectra} 

\end{figure*}

\label{sec:lines}

%{\bf It is hard to see the emission lines. Some of the emission lines not in Table 1 are also visible in the spectra (double check these lines). Usually we show only the telluric-corrected spectra. I wish to see the spectral lines in detail.}  }

%Our data for the first time zoom in four molecular clumps, IC 443 A, B, C, and
%G, in the southern molecular ridge in part of $H$ band and $K_{\rm s}$ band with high
%angular and medium spectral resolution. 

In Figure~\ref{fig:spectra}, we present spectra averaged spatially in all four regions.
All regions show strong line emission with a total of 23 detected lines and
almost no continuum emission. The background grey lines are the atmospheric
transmission at Paranal with PWV=3.5 obtained from ESO {\sc
SkyCalc}\footnote{\url{https://www.eso.org/observing/etc/bin/gen/form?INS.MODE=swspectr+INS.NAME=SKYCALC}}.
The spectrum of region G seems to exhibit more line features at the $H$-band,
which are dominated by the residuals of OH lines from the atmospheric features.
In Figure~\ref{fig:radiance}, we plot the atmospheric radiance with PWV=3.5 obtained
from ESO {\sc SkyCalc} \citep[refer to][for the OH line
catalog]{2000A&A...354.1134R}.  Table~\ref{tab:lines} lists the intensity of
the detected lines in these regions.  Although the lines between 1.80\,\mum and
1.96\,\mum seem to be well detected (Figure~\ref{fig:spectra}), especially on
the bright structures, their fluxes are highly biased by the imperfect
atmospheric corrections. Though these lines are listed in
Table~\ref{tab:lines}, we do not use them in the discussion. The 1-0 S(10) and S(11) lines are too weak in the spatially averaged spectra. However, they are detected in the bright \ce{H2} knots in regions A, C, and G (e.g. knots mentioned in Section~\ref{sec:H2} and listed in Table~\ref{tab:H2lines}), so we still list them in Table~\ref{tab:lines} as detected lines.

Most of the near-infrared emission in these molecular clumps comes from the
ro-vibrational transitions of shock disturbed warm molecular hydrogen. By
cross-checking our observed emission lines with the list of \ce{H2} lines
compiled by \cite{2019A&A...630A..58R}, we identified a total of 20 \ce{H2}
ro-vibrational lines with a signal-to-noise ratio (S/N) higher than five,
including 1-0 S(0) through S(11), 2-1 S(0) through S(5), and transitions of 3-2
S(3) and S(5), in these four regions. Among these lines, the 1-0~S(4),
1-0~S(5), and 2-1~S(4), 2-1~S(5) transitions seem to be attenuated by
atmospheric absorptions. Although the 1-0~S(3) line lies in a narrow
atmospheric window, which means that its flux might be able to be corrected, we
find still large uncertainties in the correction, especially in region~A
(Section~\ref{sec:H2}). Therefore, we ignore these five lines in the following
analysis, because of the large uncertainty in their fluxes. 

%We ignore these four lines in the
%later sections, though they are marked in Figure~\ref{fig:spectra} and listed
%in Table~\ref{tab:lines} and Table~\ref{tab:H2lines}. 

Near-infrared recombination lines of hydrogen -- Pa$\alpha$ and Br$\gamma$ (and
a few higher transitions in the Br series, up to Br10) -- are covered by
the {\it HK} grating. Unfortunately, all these lines are attenuated by the
atmospheric absorption, except for Br$\gamma$ at 2.1661\,$\mu$m, which was only
detected in region A (see details in Section~\ref{sec:Absence}).

Similarly, strong [Fe {\sc ii}] 1.6440 $\mu$m line and [Fe {\sc ii}] 1.8099
$\mu$m line are detected in region A (see Section~\ref{sec:Fe}). Nonetheless,
[Fe {\sc ii}] 1.8099 $\mu$m is attenuated by the atmosphere features.  In
regions B, C, and G, the [Fe {\sc ii}] 1.6440 $\mu$m lines are detected, while
the [Fe {\sc ii}] 1.8099 $\mu$m line is only marginal detected in region C and
shows non-detections in region B and G.  

% \usepackage{multirow}
\begin{table*}
    \centering
    \caption{Detected lines from stacked spectra of IC 443 A, B, C, and G (uncorrected for extinction).}
    \label{tab:lines}
    \begin{tabular}{lllllll}
    \hline
    \multirow{2}{*}{Transition}        & \multirow{2}{*}{$\lambda$ ($\mu$m)} & \multirow{2}{*}{$E_{\rm u}/k_{\rm B}$ (K)} & \multicolumn{4}{l}{Line Intensity ($10^{-17}$\,erg\,s$^{-1}$\,cm$^{-2}$\,sec$^{-2}$)} \\
    
                                       &                                     &                          & A                   & B                   & C                   & G                   \\
    \hline
    1-0 S(11)$\dagger$ & 1.6504 & 18979.1  & ...           & ...           & ...           & ...                 \\
    1-0 S(10)$\dagger$ & 1.6665 & 17310.8  & ...           & ...           & ...           & ...                 \\
    1-0 S(9)           & 1.6877 & 15721.5  & $16.6\pm1.2$  & $6.6\pm2.0$   & $23.8\pm1.3$  & ...                 \\
    1-0 S(8)           & 1.7147 & 14220.5  & $14.4\pm0.9$  & $5.2\pm1.6$   & $15.5\pm0.4$  & ...                 \\
    1-0 S(7)           & 1.7480 & 12817.3  & $64.0\pm1.1$  & $26.0\pm0.8$  & $89.5\pm1.1$  & $69.2\pm2.9$        \\
    1-0 S(6)           & 1.7880 & 11521.1  & $30.7\pm0.9$  & $15.0\pm0.7$  & $43.9\pm1.2$  & $49.9\pm1.2$        \\
    1-0 S(5)*          & 1.8358 & 10341.2  & $3.7\pm0.5$   & $5.0\pm0.2$   & $7.2\pm0.3$   & ...                 \\
    1-0 S(4)*          & 1.8919 & 9286.4   & $4.2\pm0.4$   & $5.3\pm0.4$   & $7.8\pm0.8$   & ...                 \\
    1-0 S(3)*          & 1.9576 & 8364.9   & $297.9\pm1.1$ & $113.1\pm1.2$ & $344.9\pm0.9$ & $301.5\pm1.7$       \\
    1-0 S(2)           & 2.0338 & 7584.3   & $84.8\pm1.1$  & $43.5\pm1.5$  & $139.1\pm1.8$ & $134.3\pm2.6$       \\
    1-0 S(1)           & 2.1218 & 6951.3   & $230.0\pm0.5$ & $129.7\pm0.7$ & $401.7\pm0.8$ & $378.5\pm2.2$       \\
    1-0 S(0)           & 2.2233 & 6471.4   & $53.7\pm0.3$  & $31.6\pm0.3$  & $95.9\pm0.3$  & $95.1\pm0.4$        \\
    2-1 S(5)*          & 1.9449 & 15762.7  & $1.5\pm0.4$   & ...           & $3.0\pm0.1$   & ...                 \\
    2-1 S(4)*          & 2.0041 & 14763.5  & ...           & $11.4\pm3.7$  & $8.5\pm0.7$   & ...                 \\
    2-1 S(3)           & 2.0735 & 13890.2  & $17.2\pm0.3$  & $5.1\pm0.2$   & $23.2\pm0.1$  & $42.7\pm0.3$        \\
    2-1 S(2)           & 2.1542 & 13150.3  & $8.3\pm0.5$   & $4.5\pm0.5$   & $13.3\pm0.3$  & ...                 \\
    2-1 S(1)           & 2.2477 & 12550.0  & $27.9\pm0.5$  & $14.9\pm0.6$  & $41.9\pm0.3$  & $37.6\pm0.8$        \\
    2-1 S(0)           & 2.3556 & 12094.9  & $7.6\pm0.8$   & ...           & $13.3\pm0.5$  & $8.6\pm1.7$         \\
    3-2 S(5)           & 2.0656 & 20855.7  & ...           & $2.4\pm0.4$   & ...           & $4.4\pm0.2$         \\
    3-2 S(3)           & 2.2014 & 19085.8  & $6.0\pm0.2$   & $5.5\pm0.8$   & $10.5\pm0.3$  & $5.8\pm1.6$         \\
    \hline
    {[}Fe {\sc ii}{]}  & 1.6440 & 11446.0  & $56.0\pm1.0$  & $11.0\pm2.0$  & ...           & $16.9\pm3.8$        \\
    {[}Fe {\sc ii}{]}* & 1.8099 & 11446.0  & $8.7\pm2.5$   & ...           & ...           & ...                 \\
    \hline
    Br$\gamma$         & 2.1661 & 154582.8 & $1.0\pm0.2$   & $<0.9$        & $<1.0$        & $<0.8$       \\
    \hline    
    \multicolumn{7}{l}{$\dagger$ Though not be listed here, these lines are detected (e.g. Table~\ref{tab:H2lines}) and marked in Figure~\ref{fig:spectra}.}\\
    \multicolumn{7}{l}{$*$ These lines are attenuated by the atmospheric absorption. Their intensities are likely underestimated.}\\
    \end{tabular}
\end{table*}

\subsubsection{\ce{H2} Rotational--vibrational  lines}
\label{sec:H2}

In total, twenty \ce{H2} lines were detected in IC 443 A, B, C, and G, ranging
from 1.6504 $\mu$m to 2.3556$\mu$m. These lines cover a wide range of the upper
level energy $E_{\rm u}$ from 6471.4 K to 20855.7 K. 

The four transitions of \ce{H2}, from 1-0 S(0) to 1-0 S(3), are the four
strongest lines at the $K_{\rm s}$ band. In regions B, C, and D, the 1-0 S(1)
transition, which has an upper level energy $E_{\rm u} = 6951.3$\,K, is the
strongest line. In region A, however, the 1-0 S(3) transition ($E_{\rm
u}=8364.9$\,K) is anomalously stronger than that of 1-0 S(1)  (see
Figure~\ref{fig:spectra} {\it top} panel). 

This anomaly leads to the ${\rm v}=1$, ${\rm J}=5$ energy level corresponding to 1-0 S(3)
transition is over-populated in region A than levels with either higher or
lower energy in the population diagram (see Section \ref{sec:popdia}).  In
fact, the ${\rm v}=1$, ${\rm J}=5$ level appears to be over-populated in all regions with
different levels, while the ${\rm v}=1$, ${\rm J}=1$, $2$, and $3$ levels are almost
linearly arranged on the population diagrams.  The 1-0 S(3) line lays inside
the range of the 1.87\,$\mu$m \ce{H2O} absorption, with high uncertainties in
the correction.  Therefore, this anomaly is more likely the result of the
errors of telluric correction and we exclude this line in the discussions
hereafter.

%On the other hand, region A is spatially close to the pulsar wind nebula, and such an over-populated could be a consequence of the radiative pumping by the UV radiation from the PWV.

\begin{figure*}
	\includegraphics[width=2\columnwidth]{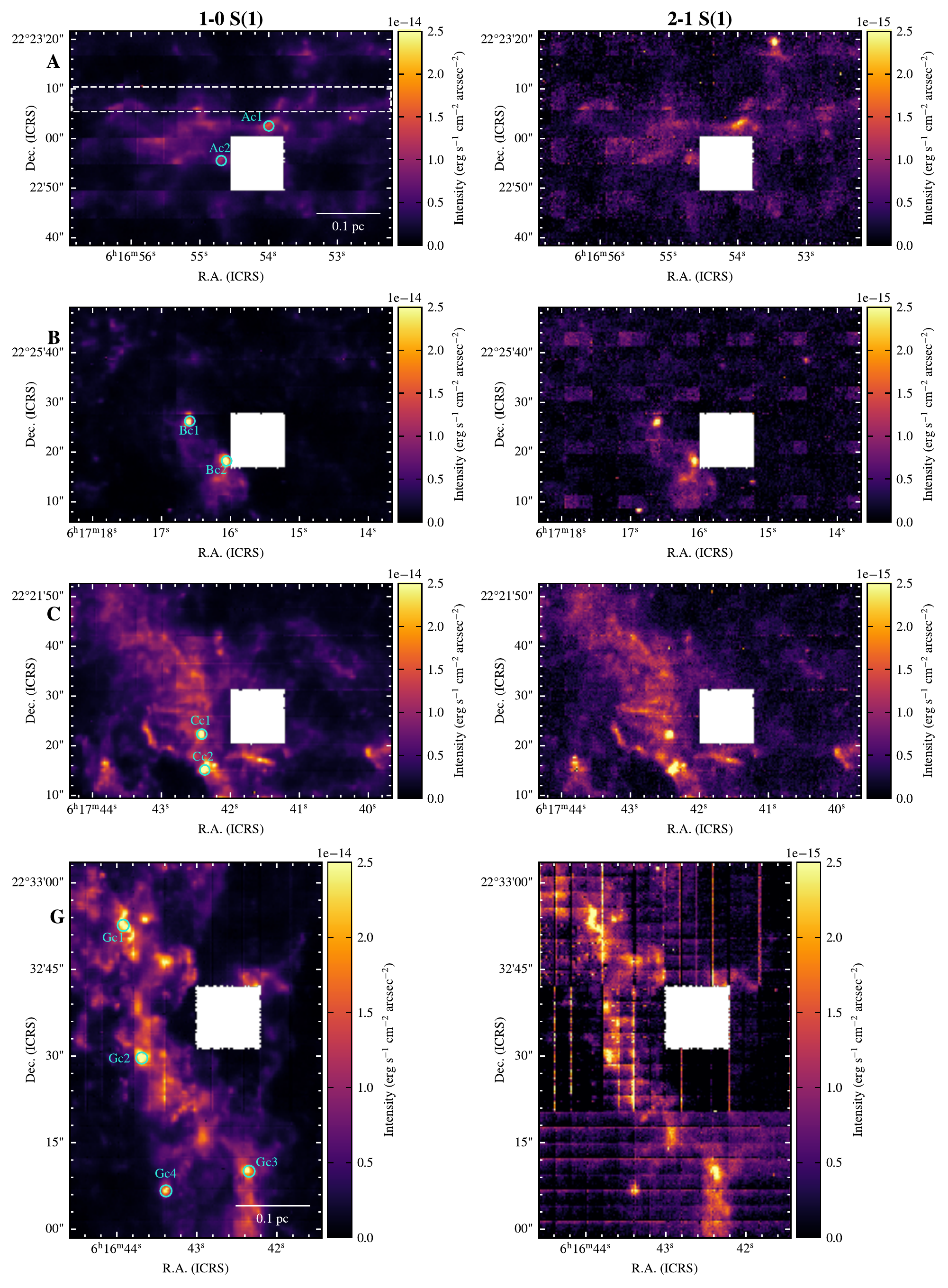}
    \caption{Spatial distribution maps of \ce{H2} 1-0 S(1) ({\it left} column)
    and \ce{H2} 2-1 S(1) ({\it right} column) emission in regions A, B, C, and
    G ({\it top} to {\it bottom}), respectively. The maximum value of the
    colourscale of 2-1 S(1) maps is $10\times$ that of the 1-0 S(1) maps.}
    \label{fig:H2Emission}
\end{figure*}

All \ce{H2} transitions show similar spatial distributions in the integrated
flux maps. In Figure~\ref{fig:H2Emission}, we present spatial distributions of the
two most often used near-infrared \ce{H2} emission line, \ce{H2} 1-0 S(1) and
\ce{H2} 2-1 S(1) for comparison. These lines are distributed across most
regions with filamentary structures on scales less than one arcmin. In all four
regions, several small and bright \ce{H2} emission clumps (hereafter
{\it knots}) appear with angular sizes of 1$''$--3$''$ ($\sim$0.008-0.024 pc)
in diameter. 

These knots are all embedded inside extended filamentary and flocculent
structures. We label a few representative ones as Ac1, Ac2, Bc1, Bc2, Cc1, Cc2,
Gc1, Gc2, Gc3, and Gc4 (see Figure~\ref{fig:H2Emission}).  Their positions, and temperature and column density derived by the strongest three \ce{H2} transitions (Section~\ref{sec:popdia}) are
listed in Table~\ref{tab:clump_parameters}.  In Table~\ref{tab:H2lines}, we list the
\ce{H2} emission lines detected in these ten small bright \ce{H2} clumps. 

\begin{table}
    \centering
     \caption{Positions and physical parameters of the ten knots.}
     \label{tab:clump_parameters}
     \begin{tabular}{lllll}
    \hline
         & R.A. (ICRS) & Dec. (ICRS)   & $T$ (K) & $\lg{N[\text{cm}^{-2}]}$ \\
    \hline
     Ac1  & 06:16:54.00  & 22:23:02.30   & $1753\pm281$        & $17.75\pm0.28$ \\
     Ac2  & 06:16:54.69  & 22:22:55.55  & $1874\pm214$        & $17.82\pm0.19$ \\
     Bc1  & 06:17:16.60  & 22:25:26.00  & $1835\pm109$        & $18.55\pm0.10$ \\
     Bc2  & 06:17:16.05  & 22:25:18.25  & $1967\pm74$         & $18.63\pm0.06$ \\
     Cc1  & 06:17:42.41  & 22:21:22.35  & $1922\pm119$        & $18.60\pm0.10$ \\
     Cc2  & 06:17:42.37  & 22:21:15.15  & $2524\pm113$        & $18.48\pm0.05$ \\
     Gc1  & 06:16:43.93  & 22:32:52.65  & $1585\pm81$         & $18.69\pm0.10$ \\
     Gc2  & 06:16:43.68  & 22:32:29.65  & $1807\pm79$         & $18.54\pm0.07$ \\
     Gc3  & 06:16:42.37  & 22:32:10.05  & $1834\pm138$        & $18.46\pm0.12$ \\
     Gc4  & 06:16:43.39  & 22:32:06.65 & $1846\pm24$         & $18.31\pm0.02$ \\

    \hline
     \end{tabular}
\end{table}

\subsubsection{The marginal detection of the Br$\gamma$ line in region A} \label{sec:Absence}

Five hydrogen recombination transitions in the Brackett series, Br$\gamma$
through Br10, lay in our wavelength coverage. However, the strongest one,
2.1661\,$\mu$m Br$\gamma$, is only detected in region A, which also coincides
with the strongest H$\alpha$ emission (see Figure~\ref{fig:ABCG}). 

The Br$\gamma$ line can hardly be recognized from any single pixel and there
are no Br$\gamma$ emission structures can be visually found in region A. To
increase sensitivity, we stack all data in region A. We estimate the flux
uncertainty with 20 nearby line-free channels. The resulted Br$\gamma$
intensity is
$1.0\pm{0.2}\times10^{-17}$\,erg\,s$^{-1}$\,cm$^{-2}$\,arcsec$^{-2}$, which is
one-ninth of the upper limit of Br$\gamma$ estimated by
\cite{1988MNRAS.231..617B} with UKIRT/CVF observations. 

\begin{figure}
	\includegraphics[width=\columnwidth]{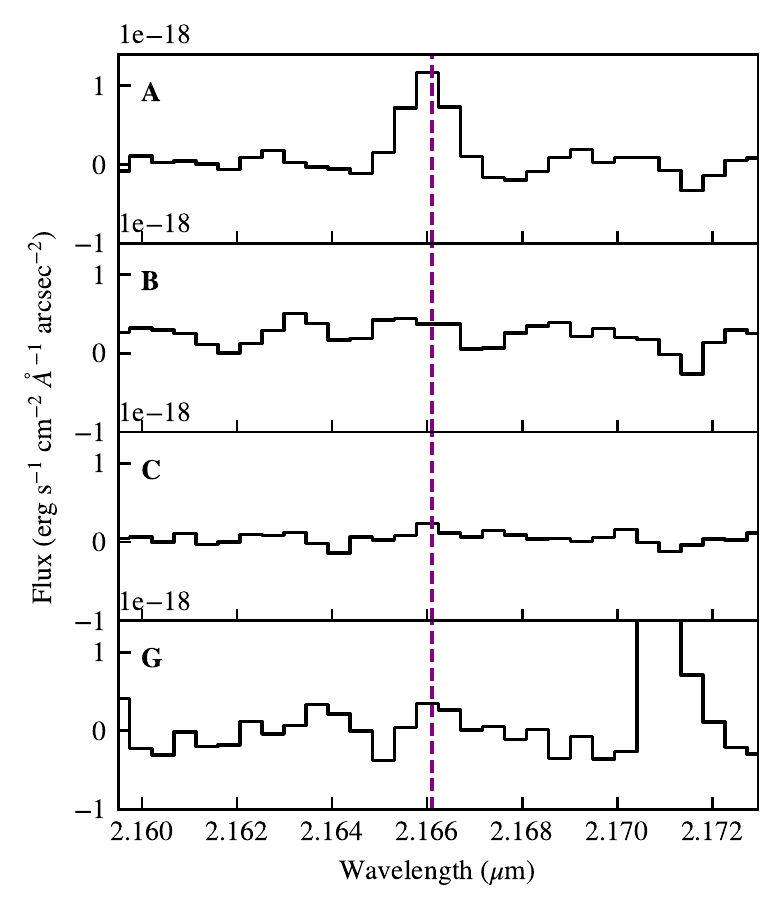}
    \caption{Stacked Br$\gamma$  spectra at regions A, B, C, and G. The purple
    dashed vertical line indicates where the 2.1661\,$\mu$m Br$\gamma$ line is
    expected.} \label{fig:Brgamma}
\end{figure}

Similarly, we also stack Br$\gamma$ spectra at regions B, C, and G,
respectively, where we only found non-detections. We thus estimate a 3-$\sigma$
upper limit of $1.0\times10^{-17}$\,erg\,s$^{-1}$\,cm$^{-2}$\,arcsec$^{-2}$ for
these three regions. In Figure~\ref{fig:Brgamma}, we present the stacked Br$\gamma$
spectra at regions A, B, C, and G. Though H$\alpha$ emission is detected in
all these regions, the Br$\gamma$ emission is absent in regions B, C, and G.

\subsubsection{[Fe {\sc ii}] emission in IC 443 A} \label{sec:Fe}
    
[Fe {\sc ii}] 1.6440 $\mu$m and 1.8099 $\mu$m  are the only metallic lines
detected in our observations. These two lines share the same upper energy level
\citep[$a_4D_{7/2}$, ][]{2016JKAS...49..109K}. In principle, we could derive
the foreground dust extinction using their line ratios. However, the 1.8099
$\mu$m line is heavily attenuated by the atmospheric absorption. Therefore, we
focus the 1.6440 $\mu$m line.

We detect the [Fe {\sc ii}] 1.6440 $\mu$m line in all four regions. Among them,
the [Fe {\sc ii}] line is strongest in region A, where diffuse filamentary
structures can be roughly recognised in the mom-0 map.  As shown in
Figure~\ref{fig:A_para}, the [Fe {\sc ii}] 1.644 $\mu$m line is about one order
of magnitude fainter than the strong \ce{H2} 1-0 S(1) line.

%The flux ratio of [Fe {\sc ii}] 1.644 $\mu$m line averaged in our four regionsmis roughly $F_A:F_B:F_C:F_G = 11.7:1.3:1:4.5$.
    
%Narrow band imaging studies have found that [Fe {\sc ii}] emission distributes in a large area of IC 443, shown as filamentary structures in the northeastern ionic shock shell, in the southern molecular shock shell and the central region inside the shells \citep[see][]{2013ApJ...768L...8K}. 

In Figure~\ref{fig:A_para}, we present the spatial distributions of [Fe {\sc ii}] and
\ce{H2} in region A, where the two lines seem roughly overlapped with each
other, although some regions reveal an anti-correlation between the two lines.
This is not consistent with the findings of \cite{2020ApJ...899...49K}, who
found that the spatial distribution of [Fe {\sc ii}] and \ce{H2} in IC 443 are
anti-correlated.  This is most likely because in region A the shock propagates
along the line of sight and the [Fe {\sc ii}] and \ce{H2} emission regions are
stratifying as successive layers on the line of sight.
    
In Figure~\ref{fig:FeVsH2}, we present the observed line profiles of Fe {\sc ii}
1.6440\,$\mu$m line (blue) and \ce{H2} 1-0 S(1) 2.1218\,$\mu$m line (red)
extracted from a horizontal stripe (Figure~\ref{fig:A_para}, white stripe) where
both [Fe {\sc ii}] and \ce{H2} emission are strong. With the very limited
velocity resolution, it is difficult to study the detailed kinematic difference
between [Fe {\sc ii}] and \ce{H2} lines. However, we can still identify that
both lines show negative average velocity. The stacked [Fe {\sc ii}] line
covers a high velocity range from $-$225\,km\,s$^{-1}$ to $\sim+$250\,km\,s$^{-1}$. 
Since the velocity profile of shocked lines is not expected to be Gaussian, we
use 16\% and 84\% integral flux to describe the line profile. The line width of
[Fe {\sc ii}] 1.6440 $\mu$m is 201\,km\,s$^{-1}$, which is 1.86$\times$ that of
\ce{H2} 1-0 S(1) (108\,km\,s$^{-1}$).

\subsection{Population diagrams} \label{sec:popdia}

The near-infrared lines of molecular hydrogen are optically thin (discussed in
Section~\ref{sec:qualitative}), so the column density of \ce{H2} at each energy
level can be calculated directly from the extinction-corrected intensity of the
specific transition,

\begin{equation}
        N_{\rm u}=\frac{4\pi\lambda I_{\rm ul}}{hcA_{\rm ul}}
\end{equation}
\label{eq:col_den}

where $A_{\rm ul}$ is the Einstein A coefficient of the transition from the
upper energy level $E_{\rm u}$ to the lower energy level $E_{\rm l}$, $I_{\rm
ul}$ is the line intensity of the transition. 

Following \cite{2011ApJ...732..124S} and \cite{2007ApJ...664..890N}, based on
the results of \cite{1995ApJ...454..277R}, we adopt $A_V=13.5$
($A_{2.12\mu{\rm m}}=1.6$) for the extinction correction in regions B and C,
and $A_V=10.8$ ($A_{2.12\mu{\rm m}}=1.3$) in region G. There is no direct
measurement of extinction for region A in the literature.
\cite{2010SCPMA..53.1357Z} find no broaden \ce{CO} emission in this region, implying a low \ce{H2} column density and dust extinction.  We
therefore adopted an $A_V$ of 2.8 \citep[$(A_{2.12\mu{\rm m}}=0.34)$,][]{1980ApJ...242.1023F},
which is
the average value of six points spreading over the optical filaments similar to those in region~A measured with the Balmer decrement.  We adopt an optical total-to-selective
extinction ratio, $R_V = A_V/E({B}-{V}) =  3.1$, and correct the
extinction using the algorithm developed by \cite{1989ApJ...345..245C}. The
wavelength-integrated flux (moment-0) maps can therefore be converted to the
column density distribution of \ce{H2} at each particular energy level.

%\zy{Does this mean the H2 column density at the specific energy level?}

%the we adopt the same $A_V=13.5$ as regions B and C, because all these regions are nearby in the southern molecular ridge. 

In Figure~\ref{fig:Population Diagram}, we present population diagrams of \ce{H2} in
ten knots, from Ac1 to Gc4. Under LTE conditions, all \ce{H2} energy levels
share the same excitation temperature and the Maxwell-Boltzmann
distribution. The natural logarithm of the column density over statistic weight $N_{\rm u}/g_{\rm
u}$ is linearly related to the energy level $E_{\rm u}/k_{\rm B}$,

\begin{equation}
    \label{equ:lte}
    \ln \frac{N_{\rm u}}{g_{\rm u}} = \ln N -\ln Q -\frac{E_{\rm u}}{k_{\rm B} T},
\end{equation}

%   \ln \frac{N_u}{g_u} = \ln N -\ln Q -\frac{E_\rm u}{k_{\rm B}T}

where $N$ is the total column density, $Q$ is the partition function of \ce{H2}
\citep{2016A&A...595A.130P}. The \ce{H2} populations seem to roughly obey the
distribution of LTE. However, two branches of populations appear on these
diagrams for transitions with upper level energies higher than 10000\,K,
suggesting that the ${\rm v}=1$ levels are more populated than the ${\rm v}=2$ levels (see
Section~\ref{sec:comparison} for further discussions). 

Nonetheless, the population diagrams in Figure~\ref{fig:Population Diagram} also show that even in these
shocked knots, the low energy levels are still in good linearity. Particularly,
the three strongest lines (from 1-0 S(0) to S(2)), which correspond to the three
lowest energy levels (${\rm v}=1$, $J = 2$, $3$, and $4$) happen to be very linear in
most regions.

Here, we use the population diagram method to analyze the \ce{H2}
emission under the assumptions of optically thin and LTE
\citep[][]{1999ApJ...517..209G}. As Equation~\ref{equ:lte} shows, by fitting the
straight line of $\ln{N_{\rm u}/g_{\rm u}}$ versus $E_{\rm u}/k_{\rm B}$, we
can directly yield a slope as $1/T$ and an intercept as $\ln{(N/Q)}$ (total
\ce{H2}). The temperature inferred here can be regarded as the excitation
temperature for the LTE assumption. In practice, we use the chi-square analysis
method to estimate the error of the fitting parameters and propagate them to
the temperature and total column density of \ce{H2}. 

Table~\ref{tab:clump_parameters} presents physical parameters estimated inside
these knots, where $T$ and $N$ are the temperature and \ce{H2} column density.
% and \ce{H2} column density derived by fitting the three lowest energy levels,
% $(T,N)_{\text{v}=1}$ and $(T,N)_{\text{v}=2}$ are derived by fitting all energy levels
% in the ${\rm v}=1$ and the ${\rm v}=2$ levels, respectively.
Most of these knots show a temperature of about 1500--2500 K and a
column density of $10^{18}$\,cm$^{-2}$. Section~\ref{sec:maps} will show that these
knots have moderate temperatures with relatively high densities inside those
shocked molecular clumps.

\subsubsection{Spatial distributions of temperature and column density traced by warm \ce{H2}}
\label{sec:maps}

As a first-order approximation, we take the strongest three \ce{H2}
transitions, from \ce{H2} 1-0 S(0) to S(2), to map the temperature and column
density distribution with population diagrams.  These transitions have the
lowest energies among our observed ro-vibrational transitions, and their upper
levels are the most populated.  We thus fit the populations acquired by these
three lines to map the temperature and column density distribution in IC~443~A
B, C, and G.  The population diagrams in regions B, C, and G show a good
linearity. 

To further increase the signal-to-noise
ratio (S/N) for extended structures, we adopt an adaptive binning scheme outlined as follows:
1). We re-bin the moment-0 map of \ce{H2} 1-0 S(2) transition with the Voronoi tessellation binning
method developed by \cite{Cappellari2003}. This binning adaptively resamples
our data to an S/N of 15 per bin for the 1-0 S(2) transition. The noise map here is obtained locally for each pixel based on its spectrum ($\sigma_{\lambda}$). 
2). We apply the same Voronoi mesh obtained earlier to the moment-0 map of the other two transitions. 
3). To exclude very large bins on the background (whose signal is mainly contributed by a few source pixels), we set a threshold of $2\sigma_\text{background}$ and masked the bins that do not satisfy this criterion.
4). We derive the temperature and column density of the bins with at least two \ce{H2} lines.
5). We keep the bins with fitted temperature $T > 5 \sigma_T$ and abandon those with uncertainty less than $5 \sigma_T$.

The derived \ce{H2} column density can be set as a lower limit of the total
column density because shocks tend to populate more \ce{H2} molecules in the
${\rm v}=0$ energy states (see Section~\ref{sec:comparison}). 
%\zy{ this needs to be confirmed} 

%At the same time, it is good to represent the warm \ce{H2} with a temperature
%of about 2500\,K.

\begin{figure*}
	\includegraphics[width=2\columnwidth]{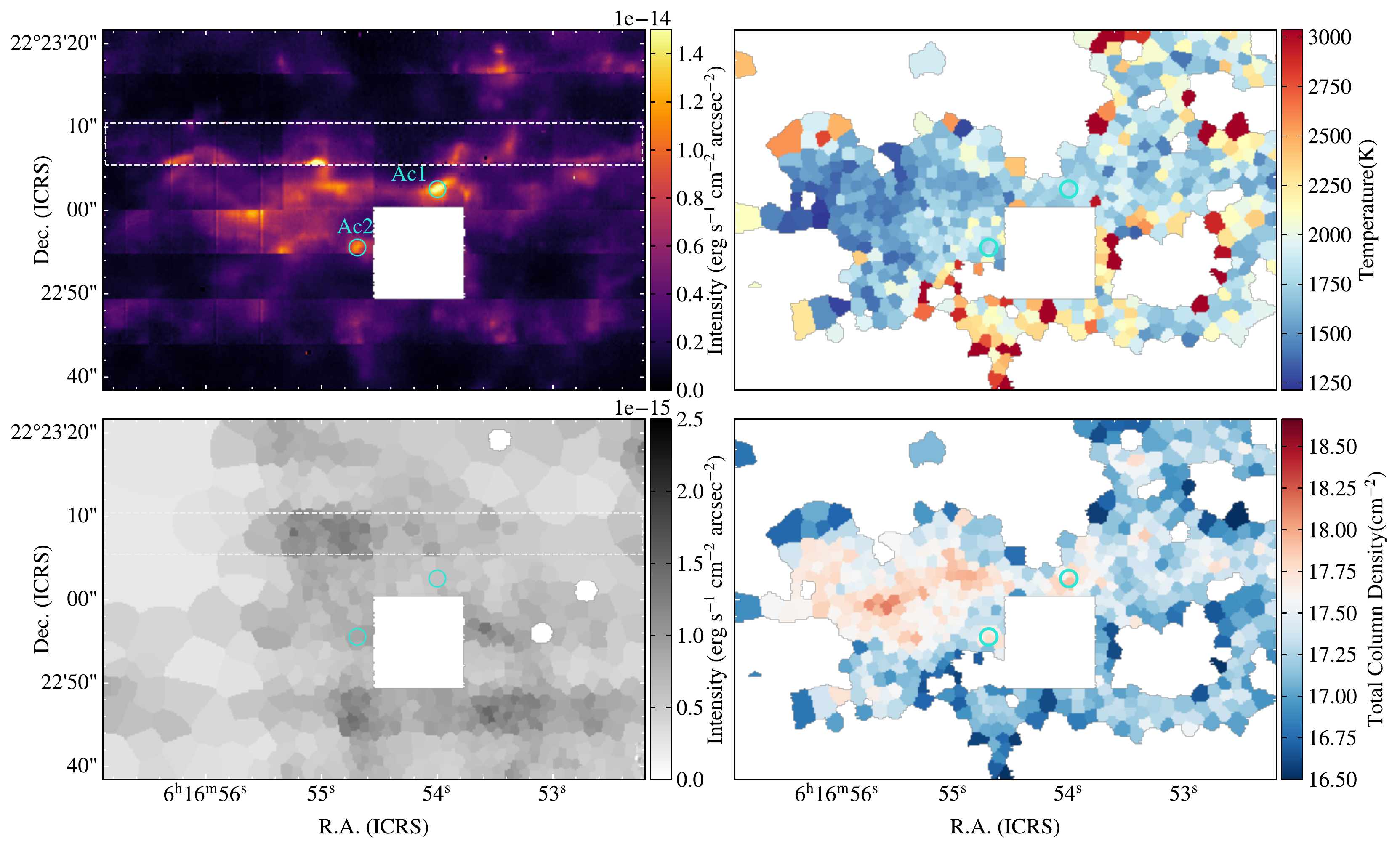}
    \caption{Maps for region A: \textit{top left:} moment-0 map for \ce{H2} 1-0 S(1) line; \textit{bottom left:} moment-0 map for [Fe {\sc ii}] 1.6440\,$\mu$m line; \textit{top right:} Temperature map with Voronoi binning; \textit{bottom right:} Column density map with Voronoi binning. Light blue circles show the position of knots Ac1 and Ac2. The white dashed line box is used to extract spectra for comparing the line profiles of \ce{H2} and [Fe {\sc ii}] (Section~\ref{sec:Fe}).}
    \label{fig:A_para}
\end{figure*}

\begin{figure}
	\includegraphics[width=\columnwidth]{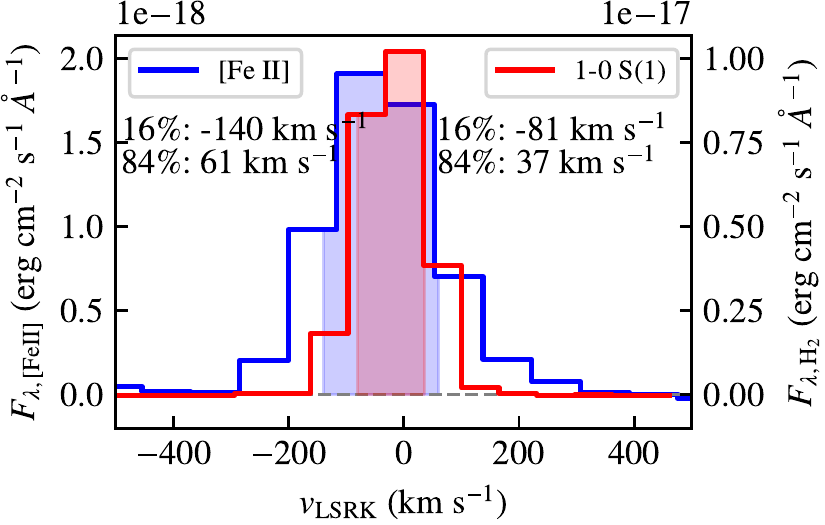}
    \caption{Observed line profiles of Fe {\sc ii} 1.6440\,$\mu$m line (blue) and \ce{H2} 1-0 S(1) 2.1218\,$\mu$m line, the 16\% to 84\% areas of the integral flux are stuffed by blue and red, respectively.}
    \label{fig:FeVsH2}
\end{figure}

\begin{figure*}
	\includegraphics[width=2\columnwidth]{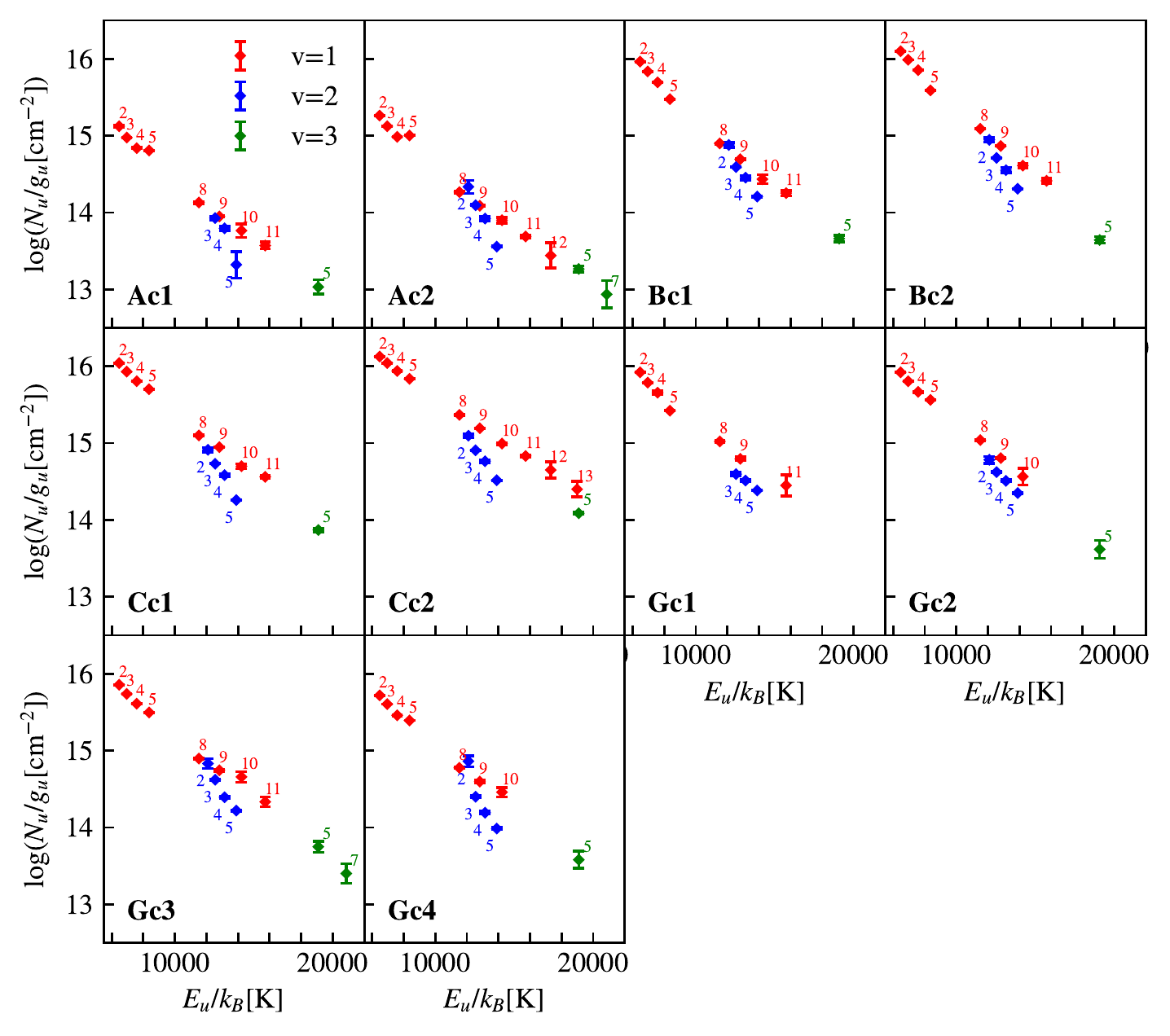}
    \caption{Population Diagrams of the ten knots. The vibrational levels are colored by red (${\rm v=1}$), blue ${\rm v=2}$, and green ${\rm v=3}$, and the rotational levels are annotated with numbers.}
    \label{fig:Population Diagram}
\end{figure*}

\begin{figure}
	\includegraphics[width=\columnwidth]{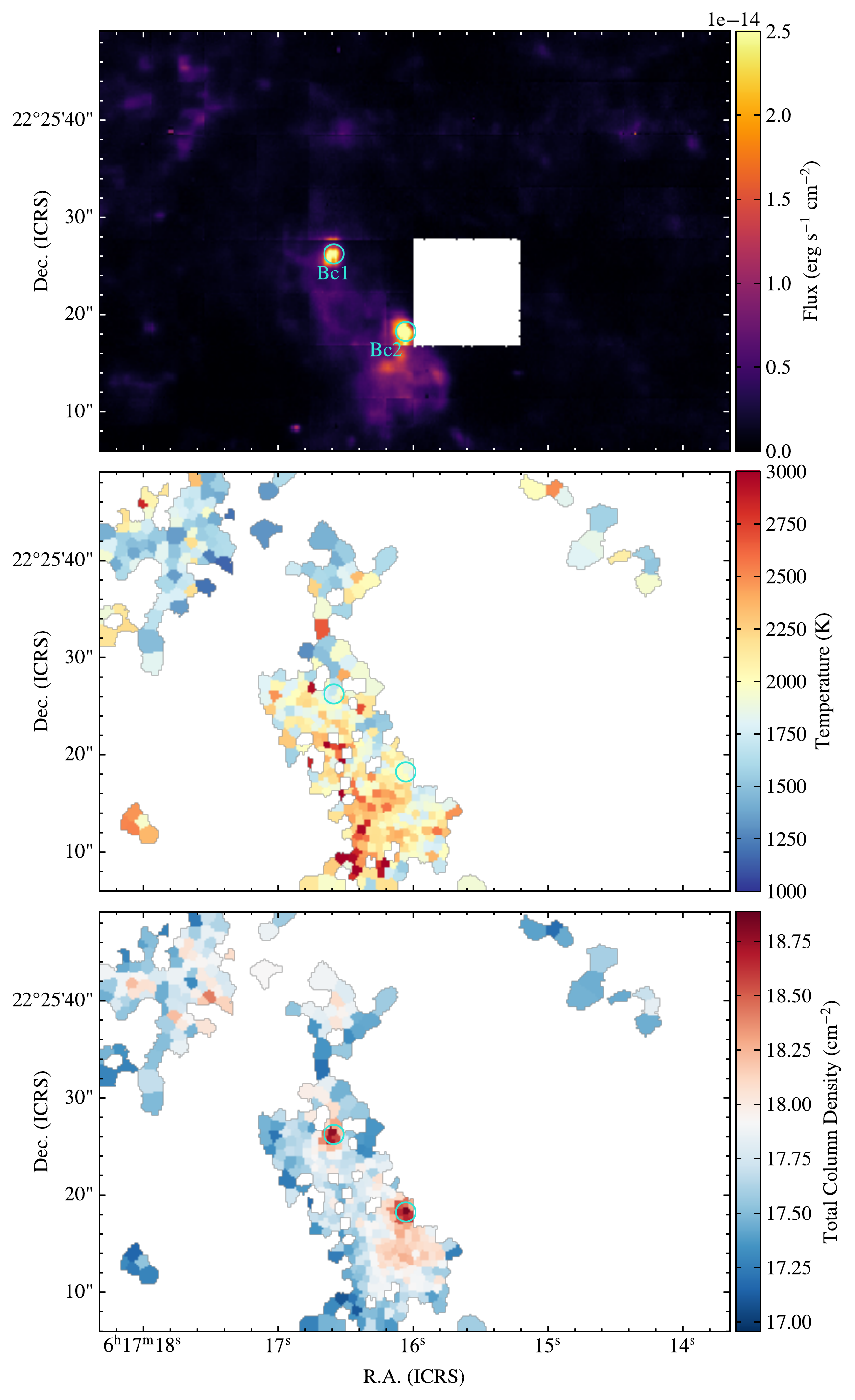}
    \caption{Maps for region B: \textit{top:} moment-0 map for \ce{H2} 1-0 S(1) line; \textit{middle:} Temperature map with Voronoi binning; \textit{bottom:} Column density map with Voronoi binning. Light blue circles show the position of knots Bc1 and Bc2.}
    \label{fig:B_para}
\end{figure}

\begin{figure}
	\includegraphics[width=\columnwidth]{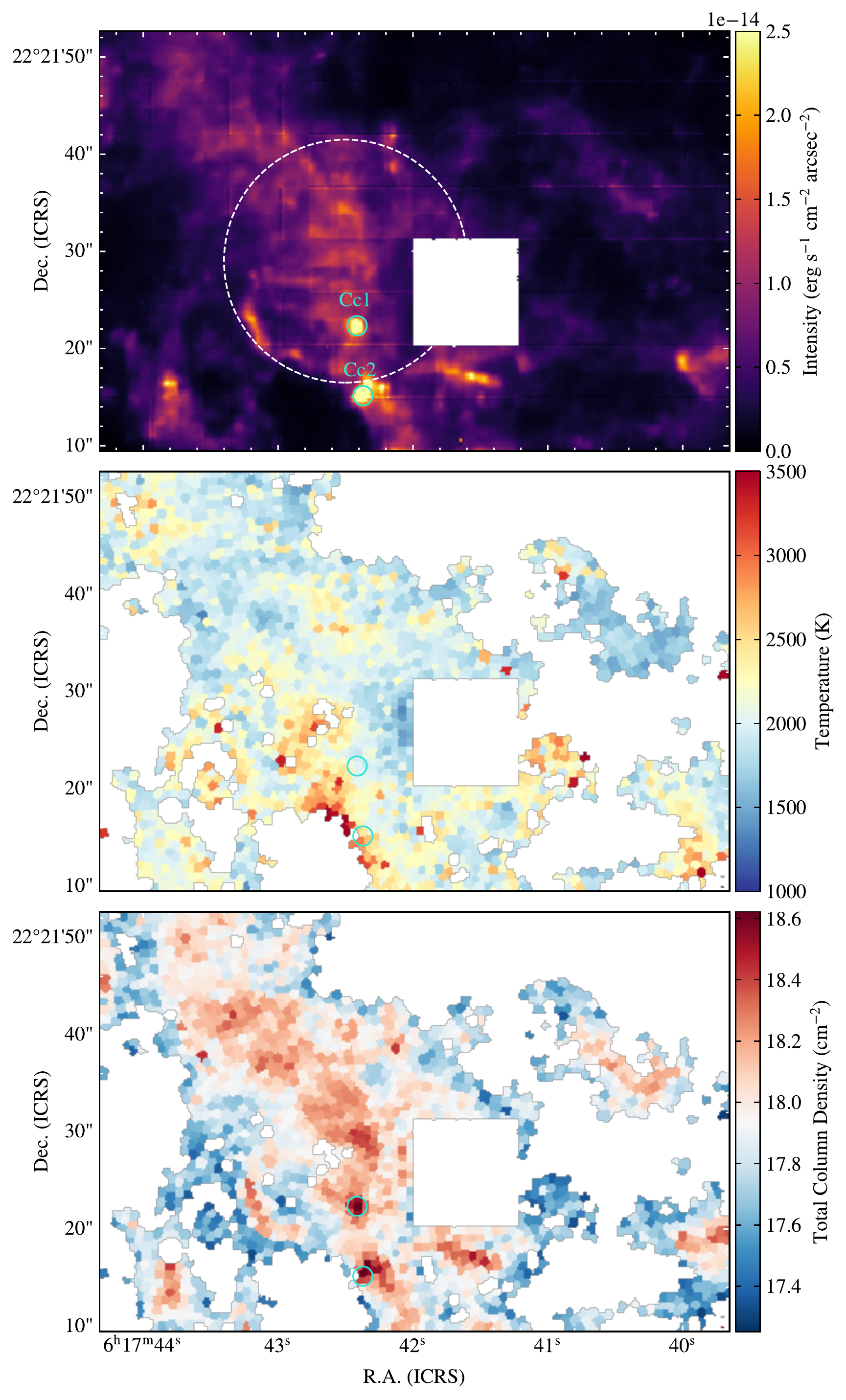}
    \caption{Same as Figure~\ref{fig:B_para}, but for region C. The white circle is the 25" HPBW of the Gaussian taper extracted in Section~\ref{sec:comparison} to compare with the \textit{Spitzer} data \citep{2007ApJ...664..890N}.}
    \label{fig:C_para}
\end{figure}

\begin{figure*}
	\includegraphics[width=2\columnwidth]{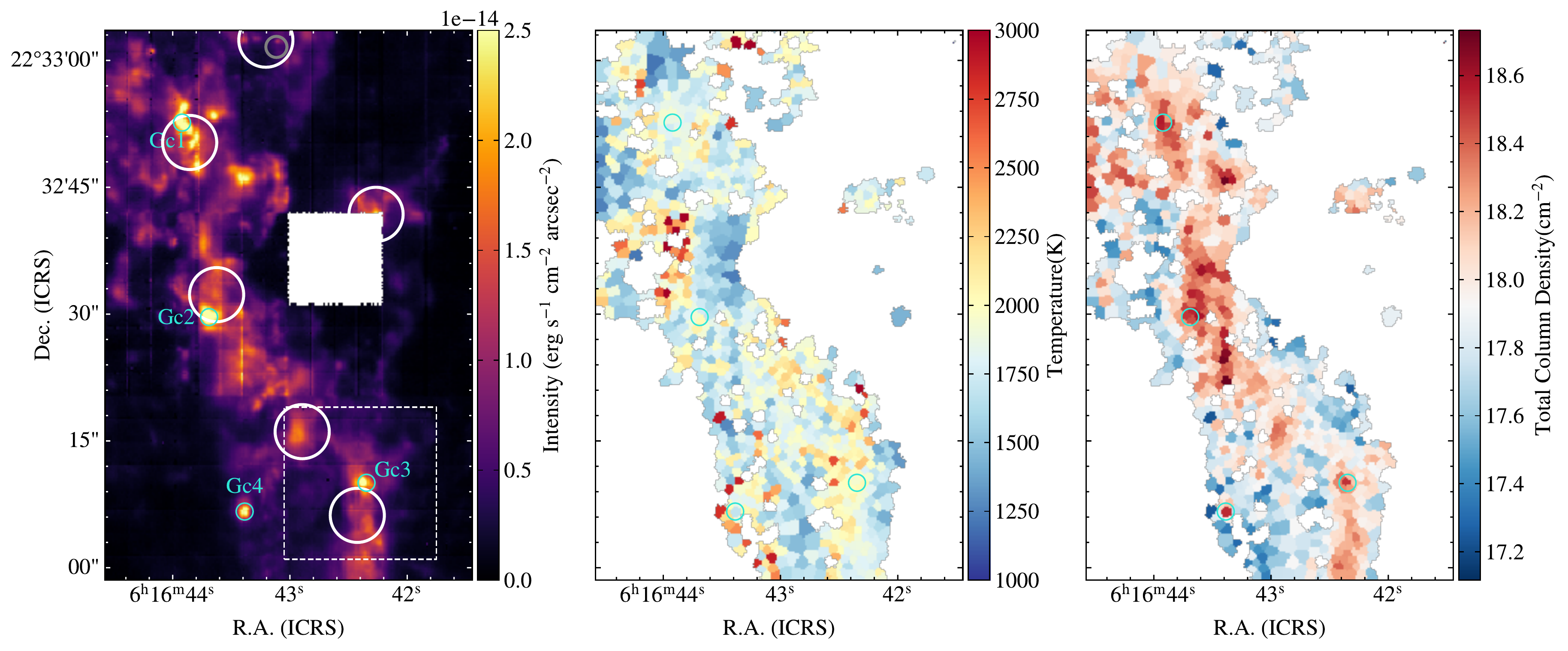}
    \caption{Same as Figure~\ref{fig:B_para}, but for region G and arranged
    horizontally. The white square is extracted in Section~\ref{sec:comparison}
    to compare with the ISOCAM data \citep{1999A&A...348..945C}. The white and
    grey circles mark the YSO candidates discussed in Section~\ref{sec:YSO}.}
    \label{fig:G_para}
\end{figure*}

\begin{figure*}
	\includegraphics[width=1.5\columnwidth]{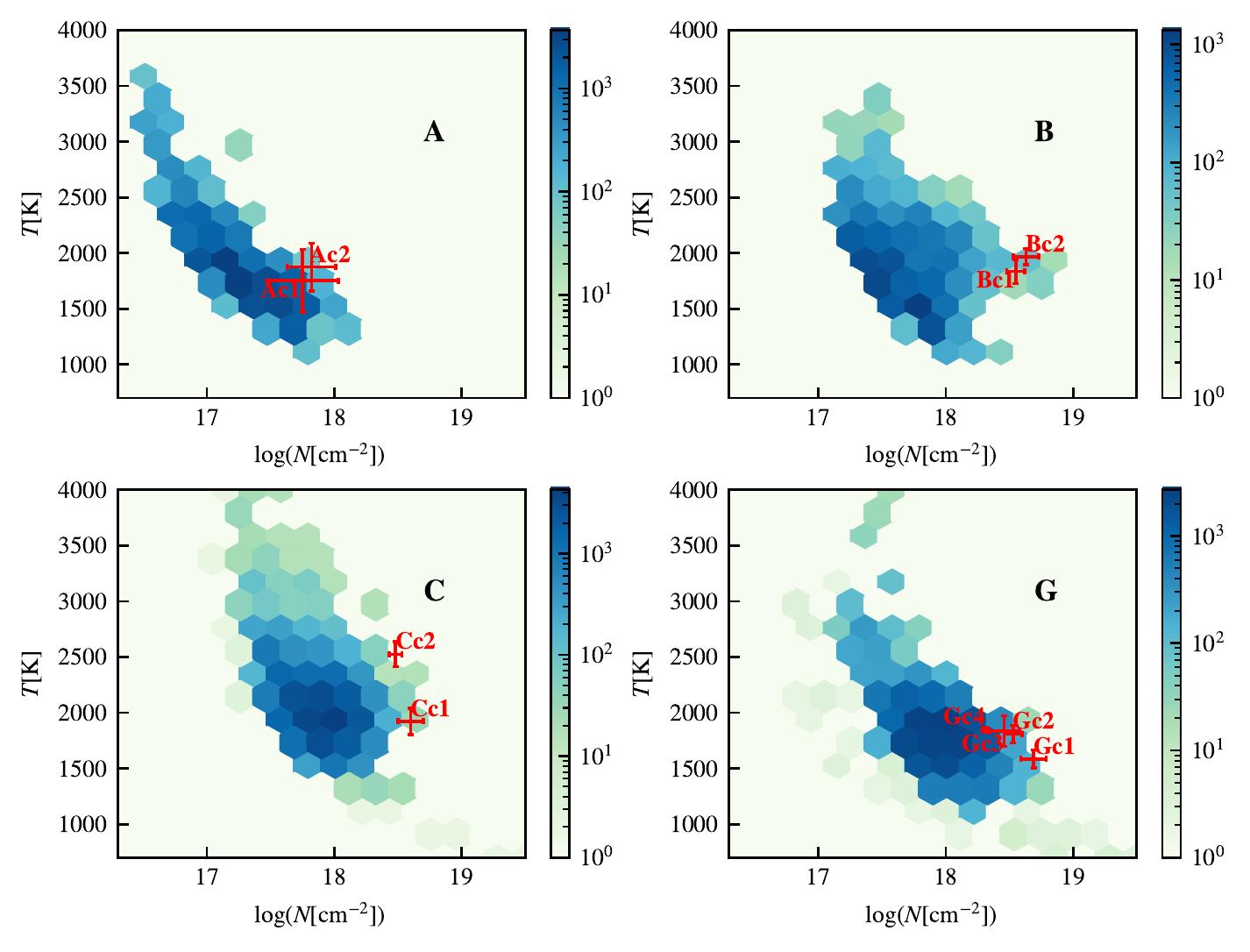}
    \caption{Phase diagrams of IC~443 A, B, C, and G derived by population
    diagrams. Ten knots, from Ac1 to Gc4, are marked with red crosses.}
    \label{fig:PhaseDiagram}
\end{figure*}

In Figures~\ref{fig:A_para}, \ref{fig:B_para}, \ref{fig:C_para}, and
\ref{fig:G_para}, we present the spatial distributions of the excitation
temperature and total \ce{H2} column density fitted from the rotation diagrams
in the A, B, C, and G regions of IC~443. Figure~\ref{fig:PhaseDiagram} presents
the phase diagrams of the excitation temperature $T_{\rm ex}$ and $N_{\rm H_2}
$ in these four regions, which all show similar ranges of $1000$--$3500$\,K and
$10^{17}$--$10^{19}$\,cm$^{-2}$, respectively. The small bright knots (from Ac1
to Gc4) are marked with 1$''$ diameter light blue cycles in
Figures~\ref{fig:A_para}--\ref{fig:G_para} and are shown as red crosses in
Figure~\ref{fig:PhaseDiagram}.

Both spatial distributions and phase diagrams present inverse correlation
between the temperature and column density in all regions, i.e. places with
high temperatures show low column density and vice versa.
Figure~\ref{fig:PhaseDiagram} shows that all those ten bright knots have
relatively high column density. 
% They can be divided into two kinds, 1) with
% high column density above $10^{18.5}$\,cm$^{-2}$ and moderate temperature about
% 1800\,K, such as Bc1, Bc2, Cc1, and knots in region G, 2) with both relatively high
% temperature and high column density, such as Ac1, Ac2, and Cc2.

The large-scale extended structure, filaments, and bright knots, which present
in the \ce{H2} emission maps, can also all be seen in the column density maps
of regions B, C, and G. In region A, however, these structures are not seen
clearly as other regions, possibly due to the relatively weaker emission. In
the following, we describe each region in detail.

%The knots we marked are all local high column density small clumps with moderately
%high temperature, which implies a hierarchical structure inside the molecular
%clumps. In the following, we describe each clump in more details.

\textbf{IC 443 A:} 
Figure~\ref{fig:A_para} shows the spatial distribution of velocity-integrated
flux of $\ce{H2}$ 1-0 S(1), temperature, and column density in IC~443 A. This region has
the lowest column density among all four fields. Most \ce{H2} gas in this
region mainly distributes along the R.A. direction, with a few branches
extending along the Dec. direction.  The south-west and north-east parts show
the highest and lowest \ce{H2} temperature, reaching $> 2500$\,K and $<
1500$\,K, respectively. The overall column density distribution presents an
anti-correlated with the \ce{H2} temperature.  Two knots, Ac1 and Ac2, have a
temperature of $\sim$ 1800\,K and a column density of $\sim 10^{17.8} \rm
cm^{-2}$.

%Brighter knots like clustered in the trunk with temperature $<3000$\,K. A
%larger knot with lower temperature and higher column density appears in the
%east of the trunk but unfortunately be sliced by the seam of the mosaic.  These
%warm diffuse structures could be the isolated rarefied small MCs or the ablated
%parts from the central large clump upon the impact of the shock.  Combining the
%relatively high temperature and low total column density, we can infer that the
%shocked molecular gas in region A is a relatively thin slab in the direction of
%the line of sight.

\textbf{IC 443 B:} 
Figure~\ref{fig:B_para} shows the spatial distributions of velocity-integrated
flux of $\ce{H2}$ 1-0 S(1), temperature, and column density in IC~443 B. The
warm \ce{H2} is mainly distributed in a clump with two bright knots, Bc1 and
Bc2, in the centre. The peak column densities of these two knots reach
$(0.5-1)\times10^{19}$\,cm$^{-2}$, while they drop sharply to $\sim
10^{17}$\,cm$^{-2}$, two orders of magnitude lower, on the edge of the whole
region.  The \ce{H2} temperature in the north part of IC~443 B is $\sim
1500$\,K, while it reaches $\sim$ 2500\,K on the south edge with low column
densities. In the north-east of region B, there appears an isolated clump with
both relatively low temperature and low column density.

%dense core with descending density layers structures, where the column density
%and descends two orders of magnitude to $10^{17}$\,cm$^{-2}$ in less than 0.1
%pc. 

%The distribution of warm \ce{H2} in region A and B implies a scenario about how
%shock waves interact with clumpy molecular clouds, we will discuss this in
%detail in Section~\ref{sec:Compare}.

%Region B is located in the centre of the $\omega$-shaped structure in the southern part of IC~443.
%The warm \ce{H2} emission is very concentrated inmore like a clump rather than a trunk. 
%The strong molecular hydrogen emission in
%IC443 B appears as a large clump with a diameter of about 20". 

\textbf{IC 443 C:} 
Region C shows the brightest \ce{H2} emission among all four regions.  As
Figure ~\ref{fig:C_para} shows, the majority of \ce{H2} emission elongates
from the northeast to the southwest, with a width $\sim$10$''$--20$''$ (0.15
pc).  Overall, the north-west region shows relatively lower temperature, and
the high temperature region concentrates in the south, likely tracing a drastic
shock heating on the edge of the gas clump. 
\ce{H2} temperature ranges from $\sim$1000 to more than 3500\,K in
temperature and from $\sim2\times10^{17}$ to $\sim3\times10^{18}$\,cm$^{-2}$.

%Flocculent structures
%are explicitly revealed inside this block as well. 

Bright knots Cc1 and Cc2 are enclosed in the trunk, showing high column
densities and temperature gradients.  Cc2 is located close to the hot peak with
more than $3500$\,K and a temperature gradient like a cooling front can be
found. The \ce{H2} emission is abrupt in front of the cooling front, which
suggests that the molecular hydrogen has not yet been disturbed and heated by
shock waves. Another possible explanation of this abruption is that there is a
sharp boundary of the molecular gas, while this did not coincide with the
extended \ce{CO} distribution (see Figure~\ref{fig:ABCG}). The temperature
decreases to $< 2000$\,K to the northwest and the transverse cooling length can
be estimated as $L_{\text{cooling}}\approx0.08$ pc ($10''$).  This also gives a
lower limit of the length of the post-shock warm zone, which agrees on the
order of magnitude with the distance of a shock front with a velocity of
$v_{\rm s}<100$\,\kms would travel in $\sim10^3$\,yr
\citep{2019ApJ...884...81R}.

In the north region of IC~443 C, the temperature decrease to less than $2000$\,K
gradually with complicated morphology, and a block of cool but high column
density gas in size of $\sim 10''$ ($\sim 0.08$ pc) appears. This cool block is
slightly fainter than the southern hot trunk because of its high column
density. The position of this cool, dense block of \ce{H2} roughly coincides
with the peak of the \ce{CO} emission, tracing a cool, low-velocity component
of the MCs (Figure~\ref{fig:ABCG}). This cool block looks like a dense bump
that divides the main emission bar into two parts.

\textbf{IC 443 G: } 
Figure~\ref{fig:G_para} shows an elongated structure laying in the north-south
direction.  Its shocked \ce{H2} column density is $\gtrsim10^{18}$\,cm$^{-2}$
and its temperature range is 1500--2500\,K, both similar to those of the other
three regions. Because the shock waves sweep the molecular gas in IC~443 G from
east to west, it seems that the warm \ce{H2} emission is encompassing the
central molecular cloud with warm \ce{H2}  shells. 

Region G has the highest concentration of bright knots (e.g. Gc1, Gc2, Gc3, and
Gc4), which accumulates in the north. All these knots have high local  column
densities of  $10^{18}\rm cm^{-2}$.  Gc1 settles in a large molecular core with
column density reaching $10^{19}$\,cm$^{-2}$, similar to Bc1 and Bc2.  Gc2 is
also located on a high column density plateau, while Gc3 and Gc4 are more
isolated.

%while the column density of the whole trunk is comparable with the dense parts in the other regions with shocks along the line of sight rather than being significantly
%larger.  This suggests the spatial distribution of warm molecular gas in region
%G is more like two thin shells encompassing a unshocked molecular pillar. 

\section{Discussion}
\label{sec:discussion}

\subsection{Physical conditions traced by \ce{H2} lines}
\label{sec:qualitative}

We first estimate the optical depth of \ce{H2} 1-0 S(1) transition, which can
be estimated as \citep[][]{2015PASP..127..266M},

\begin{equation}
    \tau_{\nu}=\frac{c^2A_{ul}}{8\pi\nu^2}[\exp(\frac{h\nu}{k_{\rm B}T})-1]\Phi_{\nu}N_u  .
\end{equation} 
\label{eq:tau}

Assuming a column density of \ce{H2} at the energy level of ${\rm v}=1$, ${\rm J}=3$ to be
$N_{{\rm v}=1,{\rm J}=2}=2\times10^{20}\text{ cm}^{-2}$ \citep[corresponding to
$N_{\text{H}_2}\approx1\times10^{22}\text{cm}^{-2}$][]{1993A&A...279..541V}, an
excitation temperature of $T_{\rm ex}=2000$\,K, a Gaussian line profile with a
full-width half maximum of $50 \text{ km s}^{-1}$ \citep{2019ApJ...884...81R},
the normalized line intensity at the peak is
$\Phi_{\nu}=1.1\times10^{-11}\text{Hz}^{-1}$. This would derive an optical
depth $\tau^{\nu}_{{\rm v}=1,{\rm J}=2} =1.2\times10^{-4}\ll 1$, which indicates that the
\ce{H2} ro-vibrational lines are optically thin.

\begin{figure*}
	\includegraphics[width=1.5\columnwidth]{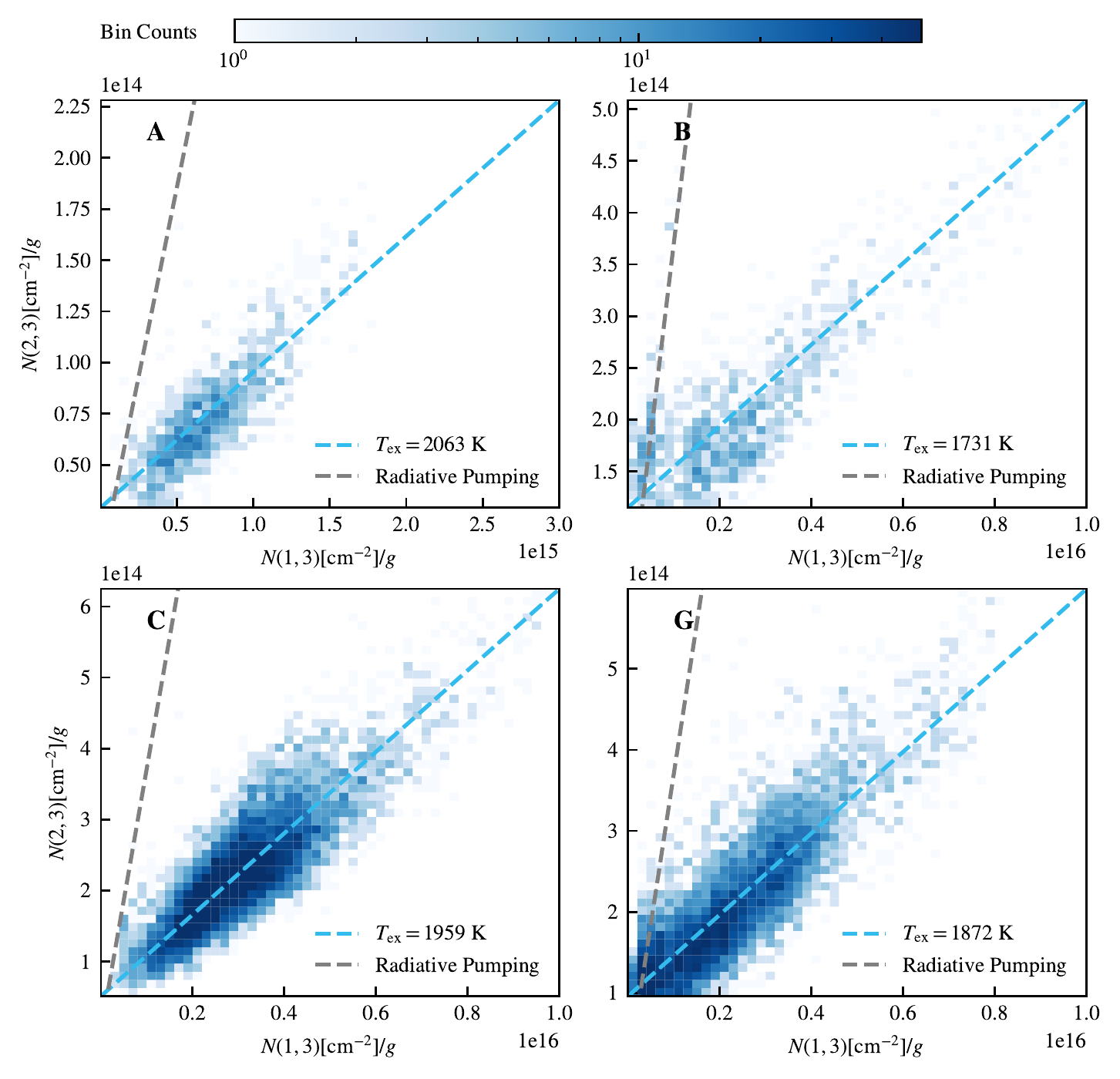}
    \caption{Two-dimensional histogram of the column density of \ce{H2} in the ${\rm v}=
    1$, $J = 3$ level versus on ${\rm v} = 2$, $J = 3$ level in the A, B, C, and G
    regions of IC~443. Blue dashed line: Predicted excitation for LTE
    conditions at the given excitation temperature; Grey dashed line: Predicted
    excitation for radiative pumping.}
    \label{fig:ddm}
\end{figure*}

The emission of \ce{H2} can be attributed to three main mechanisms: (1) being
formed in the excited state on the grain surface \citep[with a rate of
$\gamma_{\text{\ce{H2}}}=3\times10^{-17}$\,cm\,s$^{-1}$][]{1979ApJS...41..555H};
(2) radiative pumping of the \ce{H2} electronic lines followed by fluorescence;
(3) inelastic collision with H, \ce{H2}, and \ce{H+}. 

For dissociative J-type shocks, the \ce{H2} molecules can be dissociated by the
shocks but also reform rapidly in the post-shock cooling regions
\citep{1989ApJ...342..306H,2010MNRAS.406.1745F}. Indeed, the re-formation
pumping of \ce{H2} can be the predominant mechanism in the fast shocks
\citep{2011ApJ...732..124S,2012ApJ...759...34S}. However, the shocks propagated
into the prominent molecular clumps are mostly slow and non-dissociative
C-/CJ- type shocks
\citep[][]{1995ApJ...454..277R,2011ApJ...732..124S,2019ApJ...884...81R}. For the thermal collision with H atom \citep[\ce{H2}-\ce{H2} collisional dissociation has much smaller rates,][]{1967JChPh..47...54J,2000A&A...356.1010W}, the collisional dissociation rate of \ce{H2} computed by \cite{1986ApJ...311L..93D} is 
\begin{equation}
    1.0\times10^{-10}\exp{(-52000/T_{\rm n})}\text{ cm}^3\text{ s}^{-1},
\end{equation}
where $T_{\rm n}$ is the temperature of the neutral gas. Even for the gas with $T_{\rm n}\sim3000$\,K and $n_{\rm H}\sim10^4$\,cm$^{-3}$, the dissociation time scale is $\sim10^6$\,yr, which is much longer than the age of this SNR.

The non-detection of Br$\gamma$ line \citep[see Section~\ref{sec:Absence},
][]{1988MNRAS.231..617B} suggests the shock ionization is weak in these
molecular clumps. We therefore do not expect a fast ionizing shock or a strong
UV field.

Qualitatively, radiative pumping leads to cascades through excited vibrational
and rotational levels of \ce{H2}, which would efficiently form lines of the
comparable population on vibrational levels higher than ${\rm v}=1$.  Therefore, in
Figure~\ref{fig:ddm} we show the column density distributions of \ce{H2} on
${\rm v}=1$, ${\rm J}=3$ level (denote as level $j$) versus ${\rm v}=2$, ${\rm J}=3$ level (as level
$i$). The slope of this map can be used to derive the apparent excitation
temperature,

%The intensity
%ratio of 1-0 S(1) to 2-1 S(1) lines is commonly used to diagnose the excitation
%of \ce{H2}.  
%Here we equivalently plot the ``(Column) Density-Density'' map,
%two-dimensional histogram of the column density of \ce{H2} on $v=1$, ${\rm J}=3$
%level (denote as level $j$) versus on $v=2$, ${\rm J}=3$ level ($i$) in . 

\begin{equation}
\label{equ:LevelRatio}
    \frac{N_i/g_i}{N_j/g_j} = \exp{(-\frac{E_i-E_j}{k_{\rm B} T_{\text{\rm ex}}})}
\end{equation}

where $k_{\rm B}$ is the Boltzmann constant, $T_{\text{\rm ex}}$ is the apparent
excitation temperature, $N_i$, $g_i$, and $E_i$ are column density, rotational
degeneracy, and energy of \ce{H2} populated on energy level $i$, respectively.
We fit the observed data with a straight line with a least-square fitting and
plot it in blue (see Figure~\ref{fig:ddm}). 

In all four regions, the apparent excitation temperature $T_{\rm
ex}\lesssim2000$\,K. The slope of $N_ig_j/N_jg_i$ in
Equation~\ref{equ:LevelRatio} should be
$\sim1/2A_jg_j\lambda_i/A_ig_i\lambda_{\rm J}=0.37$ for the radiative pumping
dominant case, which could occur below the critical density of
$\sim10^4$\,cm$^{-3}$\citep{1989ApJ...338..197S}. This slope is much larger
than the fitting result ($<0.06$) in these four regions.  The grey lines with a
slope of 0.37 are plotted as an approximation of the radiative pumping dominant
excitation process. Thus, both formation pumping and radiative pumping could be
negligible in our regions. 

In conclusion, collisional excitation is likely to be the dominant mechanism for exciting the $E_{\rm u}\lesssim10^4$\,K levels of \ce{H2} in these dense molecular clumps with non-dissociative shocks. We will re-examine this from excitation temperature in the next section. 

%Therefore, we conclude the warm molecular hydrogen in shock dominant regions
%is
%optically thin in near-infrared bands and mainly excited by collisional
%processes. We will adopt the local thermodynamic equilibrium (LTE)
%approximation to further analyze.

\subsection{Comparison with pure rotational transitions in the literature}
\label{sec:comparison}
\begin{figure*}
	\includegraphics[width=2\columnwidth]{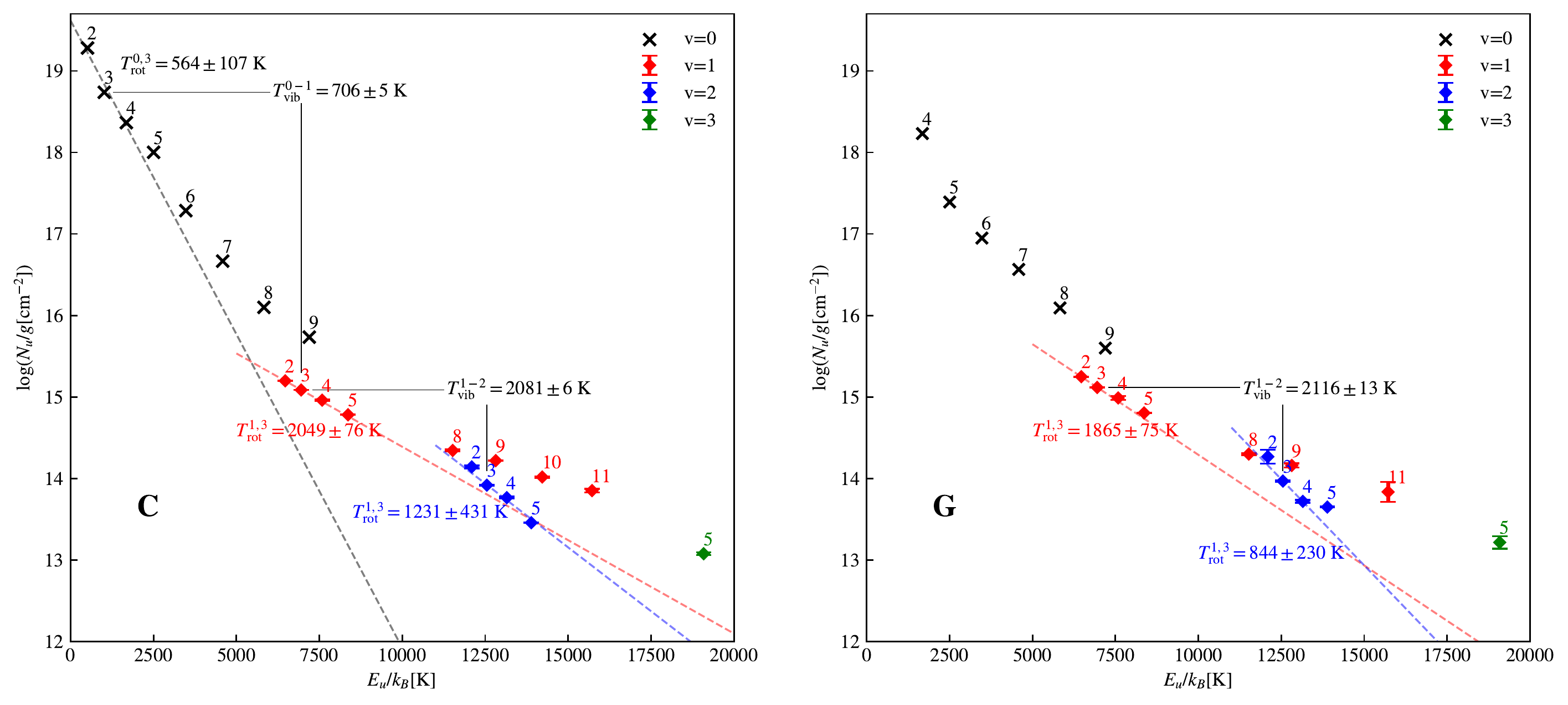}
    \caption{Population diagrams of region C and G, in comparison with previous
            mid-infrared data of pure rotational lines (black ``x''), for
            region C (\textit{left} panel) and G (\textit{right} panel). Black,
    red, blue dashed lines and texts are the fitting results for ${\rm v}=0$, $1$,
    $2$ levels, respectively.} \label{fig:comparison}

\end{figure*}

The pure rotational \ce{H2} transitions have lower energy levels than the
ro-vibrational transitions detected with VLT/KMOS. Mid-infrared observations
based on space observatories, e.g., \cite{2007ApJ...664..890N} and
\cite{1999A&A...348..945C}, have detected \ce{H2} (from 0-0 S(0) to S(7)) in
region C with \textit{Spitzer} IRS and \ce{H2} (from 0-0 S(2) to S(7)) in
region G with the Infrared Space Observatory (ISO)CAM, respectively.  Here, we
combine our new data with these mid-infrared data in regions C and G and plot
population diagrams (Figure~\ref{fig:comparison}) with a very wide energy range
from 509.9\,K to 19085.8\,K.

First, we correct the mid-infrared data with the same extinction adopted in
this work. For clump C, \cite{2007ApJ...664..890N} present a Gaussian weighted
average spectrum within a circular region centred at (06:17:42,$+$22:21:29),
with a HPBW of 25$''$. This subregion contains the molecular clump in region C
and knot Cc1, as labeled with the white circle in Figure~\ref{fig:C_para} {\it
top} panel. We extract the spectrum from our data with the same Gaussian
weighting, which is plotted in the {\it left} panel of
Figure~\ref{fig:comparison}.  For region G, the ISOCAM data is compared with
the spectrum extracted from an 18$''\times18''$ region at ``Peak B'' in
\cite{1999A&A...348..945C}, as labeled with the white square in
Figure~\ref{fig:G_para} {\it left} panel. The population diagram is present in
the {\it right} panel of Figure~\ref{fig:comparison}.

In both regions, the population diagrams have three vibrational levels, namely
${\rm v} = 0$, $1$, and $2$, respectively. Only one data point exists for ${\rm v} = 3$
data at each region, so we overplot it in the figure without any fitting. The data points in all vibrational levels follow relatively straight lines, with
different slopes. The slopes tend to flatten at high $J$ levels because of the presence of temperature gradients and multiple temperature components. 
Following the method presented in Section~\ref{sec:popdia}, we derive the $T$ and $N$ by fitting the lowest ${\rm J}=2,3,4$ levels at the ${\rm v}=1$ branch and plot the LTE population of the fitting results as the red lines in Figure~\ref{fig:comparison}. The flattening of the slope can be found by comparing the ${\rm J}=5,8,9,10,11$ levels with the red lines. 

In C-type shocks, the shock transition of temperature is continuous from $\sim10$\,K to $>10^3$\,K succeeding with radiative cooling. Such temperature gradients
result in that the hotter gas with a smaller column density dominates the excitation of high energy states, while the cooler gas with a larger column density dominates that of the low energy states. Thus, as an integration in the line of sight, the apparent excitation temperatures of the higher energy levels can be larger than that of the lower levels but their column densities are orders of magnitudes smaller. The temperature and column density derived from the low ${\rm v} = 0$ levels represent the majority of the molecular gas involved in the shock transition which perhaps includes a wide precursor, and those derived from the low ${\rm v} = 1$ levels can represent the gas in the $10^3$\,K warm shock layers (see also Section~\ref{sec:contribute}).

The temperature obtained with the above method can also be regarded as the rotational excitation temperature of ${\rm v}=1, {\rm J}=3$ level, hereafter $T_{\rm rot}^{1,3}$. Similarly, we can derive $T_{\rm rot}^{0,3}$ and $T_{\rm rot}^{2,3}$ to present the rotational excitation on a specific vibrational level. Meanwhile, the vibrational temperatures between ${\rm v}=v, {\rm J}=3$ and ${\rm v}=v', {\rm J}=3$ levels are simply derived by
\begin{equation}
    T_{\rm vib}^{v-v'} =\frac{E_{v,3}-E_{v',3}}{k_{\rm B}\ln{(N_vg_{v'}/N_{v'}g_v)}}.
\end{equation}

The derived rotational and vibrational excitation temperatures are marked in Figure~\ref{fig:comparison}. The vibrational temperatures $T_{\rm vib}^{0-1}$ and $T_{\rm vib}^{1-2}$ at the ${\rm J}=3$ levels are consistent with the rotational temperatures $T_{\rm rot}^{0,3}$ and $T_{\rm rot}^{1,3}$, respectively. This result suggests that the rotational and vibrational excitation should undergo the same mechanism, as we discussed in Section~\ref{sec:qualitative}, collisional excitation.

The population diagrams of different v levels are not fully
consistent with each other: at similar $E_{\rm u}/k_{\rm B}$, the ${\rm v}=0$ levels
are more populated than the ${\rm v}=1$ levels and the ${\rm v}=1$  levels are also more
populated than the ${\rm v}=2$ levels. These features imply that the population diagrams can not be fully reproduced by simply combining multiple LTE components with different $N$, and $T$, which could be feasible when modeling the OMC-1 \citep{2000A&A...354.1134R}. \cite{2011ApJ...732..124S} have successfully reproduced the level population of shocked \ce{H2} in IC~443~C and G using the combination of two power-law thermal admixture models which assume that the \ce{H2} column density in the C-shock is related to the temperature as ${\rm d}N=aT^{-b}{\rm d}T$. 

It is worth noting that the ${\rm v}=1$
branch shows the flattest trend which leads to the highest temperature and lowest column density among all vibrational levels, while ${\rm v}=2$ branch has only moderate excitation temperature of $\sim1000$\,K. The most reasonable explanation of the moderate temperature and column density derived from ${\rm v}=2$ branch is that we only observed the lowest four levels while the slope can also flatten rapidly at higher levels. In fact, the flattening trend of the ${\rm v}=2$ branch has appeared in the population diagram of region G (Figure~\ref{fig:comparison} {\it right}). Fitting with only the lowest transitions would underestimate the temperature and overestimate column density. 

Such ${\rm v}=2$ population diagrams can also be found in the simulated shocks \citep[e.g.][]{2007MNRAS.382..133W,2019MNRAS.489.4520N} but it was not been interpreted in the literature. A detailed investigation needs the observations of more ${\rm v}=2$ levels, whose ro-vibrational transitions, unfortunately, lie in the 1.87$\mu$m opaque atmospheric band. Future {\it James Webb Space Telescope} observations can exhibit the most complete \ce{H2} population diagrams and help the community with it.

\subsection{Warm molecular gas in shocked regions}
\subsubsection{Multi-band comparison}
\label{sec:Compare} 
%The shock waves impact on the quiescent molecular gas and the gas begin to be compressed, disturbed, and heated up. 
\begin{figure*}
	\includegraphics[width=2\columnwidth]{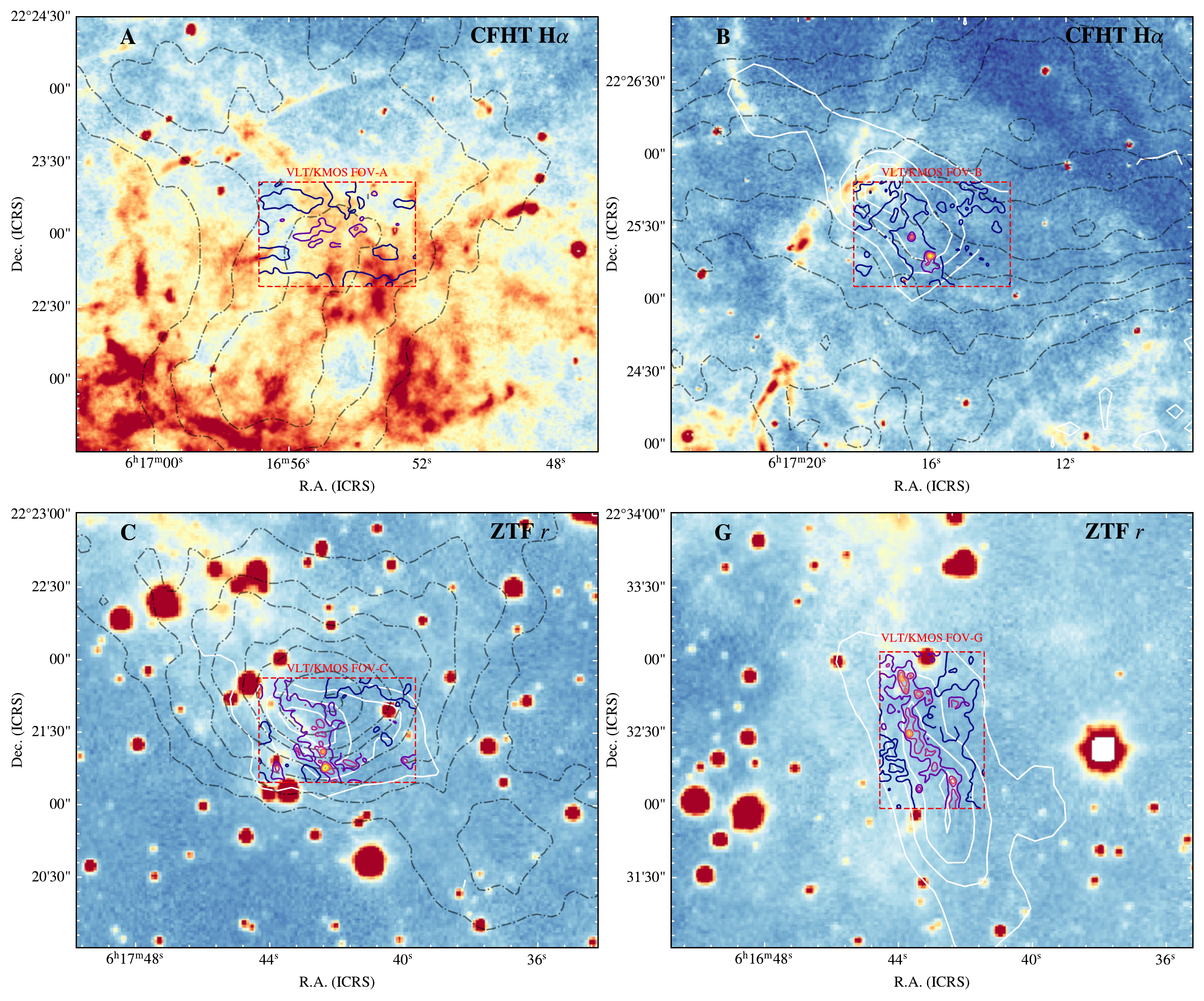}
	\caption{Comparison of the distributions of H$\alpha$, high-velocity H {\sc i}, \ce{H2}, and shocked \ce{CO} in IC 443 A, B, C,and G. Background colormap: the CFHT H$\alpha$ (region A and B) or ZTF \textit{r} band images (region C and G); Black contour: the moment-0 map of high-velocity H {\sc i} ($-100<v_{\text{LSR}}<-20$\,km\,s$^{-1}$), contour levels start from 0.7\,Jy\,km\,s$^{-1}$ and increase by 0.3\,Jy\,km\,s$^{-1}$ per level; White contour: moment-0 map of shocked CO 3-2 (both redshift and blueshift in region G), contour levels start from 100\,K\,km\,s$^{-1}$ and increase by 100\,K\,km\,s$^{-1}$ per level; purple contour: our KMOS \ce{H2} 1-0 S(1) data, contour levels start from $1.25\times10^{-15}$,erg\,s$^{-1}$\,cm$^{-2}$\,arcsec$^{-2}$ and increase by $5\times10^{-15}$,erg\,s$^{-1}$\,cm$^{-2}$\,arcsec$^{-2}$.}
    \label{fig:ABCG}
\end{figure*}

In this section, we compare the multi-gas-phase distribution and excitation in
the shock environment of IC~443, by comparing the data for our VLT/KMOS \ce{H2}
observations with multi-band ancillary data (Section~\ref{sec:ancillary})
including JCMT CO 3-2 data, VLA \& Arecibo H~{\sc i} 21-cm data, and the
CFHT/SITELLE H$\alpha$/ZTF \textit{r} band (which is dominated by the H$\alpha$
emission) photometric data. 

In Figure~\ref{fig:ABCG}, the background colormap shows the H$\alpha$/ZTF
\textit{r} band images. The black contour shows the moment-0 map of
high-velocity H{\sc i} ($-100<v_{\text{LSR}}<-20$\,km\,s$^{-1}$). The white
contour shows the moment-0 map of shocked CO $J$=3-2 (both redshifted and
blueshifted components from region G). The purple contour presents our
KMOS \ce{H2} 1-0 S(1) data. 

The \ce{H2}, high-velocity H~{\sc i}, and CO emission all show clumpy
structures. All these species co-exist in the same regions, with peaks slightly
staggered. On the other hand, the distribution of H$\alpha$ seems more extended
with filamentary structures. All \ce{H2} emission is enclosed by CO and
high-velocity H~{\sc i} emission. In A and B regions, H$\alpha$ seems to have
no spatial relation with the \ce{H2} morphology, while in C and G regions,
elongated H$\alpha$ emission seems to be aligned with the \ce{H2} structures
with slight offsets.

%The fragmented and filamented  H$\alpha$ emission structures looks like a sort
%of veil or fiber net. These homogeneity and heterogeneity of emission
%structures of different species reveals the structure of the shocks.

%An interesting phenomenon is that there are always several bright clumps
%smaller than 1" embedding in the large-scale \ce{H2} clump among our four
%FOVs. These small clumps can be divided into two categories: 1. clumps with
%high temperature in rarefied gas like Ac1, Ac2, Gc2, Gc3, and Gc4; 2. clumps
%with relative high column density but moderate temperature like Bc1, Bc2, Cc1,
%Cc2, and Gc1. Since their spatial scale is comparable to the seeing of our
%observations, the real size of such small structures may not be well-resolved.
%A possible origin of the latter ones can be the Rayleigh-Talyor instability
%(RTI), which can occur when the pressure gradient is  in the direction
%opposite to the density gradient. A MC shocked by blast waves is a reasonable
%situation for appearance of RTI and shock waves can speed up the transition to
%instability significantly \citep[][]{2006PhPl...13f2705Z}. Rarefied gas in the
%outer layer of the molecular clump can penetrate into denser inner layers
%through RTI. Once RTI occured, the flow would develop into a high Reynolds
%number turbulence with very strong nonlinearity and the mechanic energy would
%dissipate significantly by the turbulence.

\subsubsection{Stratified shock structures}
\label{sec:Stratify}

%In region A, As Figure~\ref{fig:ABCG} shows, the strongest H$\alpha$
%emission among our four FOVs is detected and looks like a knot of several
%fibers in shape. 

Region A exhibits the strongest H$\alpha$ emission among all four regions,
composed with several expanding shell-like structures.  The H {\sc i} emission
in region A is enclosed by the H$\alpha$ shell in the south.  The \ce{H2} clump
(our FOVs) is on the ridge of the H {\sc i} structure along the north-south
direction. The spatial overlapping of \ce{H$\alpha$}, H{\sc i}, \ce{H2}, and
[Fe {\sc ii}] (Section~\ref{sec:Fe}) implies that the shock waves are
propagating along the line of sight and interacting with the rarefied atomic
gas and dense clump successively. 

% Both the H$\alpha$ emission structures and H {\sc i} clump have a
% tendency to extend to the northeast. 

%Since Br$\gamma$ is not detected with a very low upper limit, the high-velocity of H {\sc i} 21-cm emission can only come from the shock accelerated atomic H or fractionally dissociated \ce{H2}. 

As Section~\ref{sec:Fe} mentioned, region A shows strongly broaden [Fe {\sc
ii}] emission from the same region of the \ce{H2} emission. 
To produce the [Fe {\sc ii}] lines, it requires a fast shock with a velocity
$>$50\,\kms to release Fe locked in the grain
\citep{1994ApJ...433..797J,1995Ap&SS.233..111D}. 
The observed broad velocity profile of [Fe {\sc ii}] is consistent with the
fast shock scenario.  Therefore, we propose that in region A the shock still
keeps its high velocity, either because the molecular clump here is
intrinsically rarefied, or because it only recently starts to interact with the
dense molecular clumps and produces \ce{H2} emission. On the contrary, the
shocks have been significantly decelerated by the dense molecular clumps in
regions B, C, and G, which all show bright \ce{H2} emission and weak [Fe {\sc
ii}] emission.

If the foreground gas is rarefied, the ISM in region A can be 
dissociated or ionized to H {\sc i} and \ce{H+}, and the grain can release Fe
into gas. However, the detection of HCO$^+$\citep{1992ApJ...400..203D}, which
has a high critical density $\sim 10^4\rm cm^{-3}$, needs the existence of
high-volume-density \ce{H2} gas.

If the foreground gas is stratified with dense cores in the centre, the fast shock
can still heat up surrounding diffuse gas, producing high-velocity H {\sc i},
\ce{H+}, and [Fe {\sc ii}] lines.  
The central core part of the MC is still
undisturbed. This scenario would expect a velocity difference among all these
lines: ${\rm v}_\text{[Fe {\sc ii}]} \sim {\rm v}_{{\rm H}\alpha} > {\rm v}_\text{H{\sc i}} > {\rm v}_\text{\ce{H2}}$. 
From Fig \ref{fig:FeVsH2}, the linewidth of [Fe {\sc ii}] is larger than that of the \ce{H2} lines. 
So we would be inclined to this case.  Future high-velocity resolution observations could
be useful to fully test these two scenarios. 
 
 %The shock wave just begins to interact with a MC with a dense core and an
 %increasing density gradient as it moves close to the core. 
% Rarefied gas surrounding the dense
% clump is apt to be heated to very high temperatures, and be dissociate to
% disturbed H {\sc i} even ionized to \ce{H+}, the grain also is sputtered and
% release Fe into gas, while the main part of the MC is still undisturbed. 
 
%Since the foreground \ce{H2} is optically thin and no CO clump is detected in
%region A, of course, we can't rule out the possibility that the molecular
%density of region~A is intrinsically low and there is actually no molecular
%cloud core.

%Even so, the observed appearance of various species overlapping at the line of
%sight implies region A could be an intriguing test field to reconstruct the
%tomogram of shocked multi-phase gas.

In regions B, C, and G, H$\alpha$ emission is much weaker than that in region
A.  The morphologies of H$\alpha$ in these regions are more extended, often
appearing only on one side of the molecular clumps.

Similar to region A, shock waves in regions B and C are also propagating along the
line of sight. Our JCMT data and \cite{2010SCPMA..53.1357Z} show a strong
$^{12}$CO clump in regions B and C, with $^{13}$CO emission, indicating the
existence of a concentration of dense molecular gas. This distribution of
atomic and molecular gas is interpreted by a schematic diagram of the molecular
clump in region B proposed by \cite{2010SCPMA..53.1357Z}, where the ionized and
atomic gas in the outer layer of molecular clouds are  accelerated and ablated
by the fast shocks.  The CO and \ce{H2} emissions mainly come from the dense
core surrounded by H {\sc i}. The temperature and column density distribution
shown in Figure~\ref{fig:B_para} also support this scenario, where the dense
gas is surrounded by the rarefied hot gas.

In region G, the shocks propagate perpendicular to the line of sight. The
\ce{H2} and CO clumps show spatially coexisting structures, while the H$\alpha$
emission only appears at the east of the molecular clump. This can also be
interpreted by a similar shock disturbance scheme as the vertical line of sight
case.

Such a scenario is also similar to a classic astrophysical hydrodynamics test
problem, the {\it blob test}
\citep[e.g.][]{2007MNRAS.380..963A,2015MNRAS.450...53H}, where a spherical,
dense, and cold cloud in the pressure equilibrium with a uniform rarefied, warm
background gas is disrupted by a supersonic wind. Hydrodynamic simulations
have shown that the ram-pressure ablates and disrupts the clouds with
a complicated mixture of the Kelvin-Helmholtz instabilities (KHIs) and the
Rayleigh-Taylor instabilities \citep[RTIs][]{1961hhs..book.....C} developing
from the front of the cloud during the stripping. The KHIs and RTIs fragment
the cloud into several small dense clumps and filaments with tails before the
cloud finally being destructed \citep[e.g. see Section 4.4.3 and Figure 24
of][]{2015MNRAS.450...53H}. 

The hierarchical structures of embedded knots and filaments we observed in the
shocked warm \ce{H2} emission regions (Section~\ref{sec:H2}) can be the
realistic astrophysical counterpart of such a disrupting cloud. The filamentary
H$\alpha$ emission, surrounding the molecular clumps, can be regarded as the
ablated filaments and tails. Under such a cloud-fragmentation scenario, the
\ce{CO} and \ce{HCO+} counterparts of the small bright \ce{H2} knots can be
tested with high-resolution sub-millimetre observations.

\subsubsection{How much does the warm \ce{H2} contribute in shocked regions?}
\label{sec:contribute}

Our detected $H$- and $K$- band \ce{H2} lines mostly correspond to the
high-volume-density component in the power-law thermal admixture model
\citep{2011ApJ...732..124S}. In the region G , we have obtained column density of the warm \ce{H2} trunk is $\gtrsim10^{18}$\,cm$^{-2}$ and hardly reaches
$10^{19}$\,cm$^{-2}$ (see Section~\ref{sec:maps}). 

In the same region G, \cite{2020A&A...644A..64D} obtained the total
column density of shocked \ce{H2} to be
$(0.33-1.44)\times10^{21}$\,cm$^{-2}$ using the standard \ce{CO}/\ce{H2}
abundance ratio ($10^{-4}$), which is plausible to be adopted even in these
shocked clumps \citep{1993A&A...279..541V}.  Comparing results from both
population diagrams of CO lines and the Large Velocity Gradient (LVG) method
with the expanding spherical geometry in the \verb"RADEX" code
\citep{2007A&A...468..627V}.  \cite{2020A&A...644A..64D} find that column
density of shocked high-velocity CO is three times more than that obtained from
population diagrams.  

This suggests that the warm
molecular hydrogen with temperature $\sim2000$\,K contributes to less than 1\%
of the total mass of the shocked \ce{H2}.

\subsection{Emission line contribution to 2MASS $H$ and $K_{\rm s}$ bands}

The Two Micron All-Sky Survey \citep[2MASS; ][]{2006AJ....131.1163S} database
has been widely used to provide the near-IR photometry of IC~443 and other SNRs
in the $JHK_{\rm s}$ bands \citep[e.g.][]{2001ApJ...547..885R, 2014ApJ...788..122S,
2003ApJ...592..299R}. Using the {\it HK} grating, our KMOS observations cover the
2MASS $H$ ($1.5-1.8$\,$\mu$m) band and $K_{\rm s}$ ($2-2.32$\,$\mu$m) band
simultaneously and therefore provide an opportunity to measure contributions
from \ce{H2} lines. 

Extended emission at $K_{\rm s}$ band is often thought to be dominated by \ce{H2} 1-0
S(1) 2.1218 $\mu$m, which is actually quite strong in our observed fields.
However, all the four strongest \ce{H2} transitions (from 1-0 S(0) to S(3)) lay
in the $K_{\rm s}$ band. In fact, 1-0 (from S(0) to S(3)), 2-1 (S(5) to S(0)), and
3-2 (from S(1) to S(5)) could contribute to the flux of $K_{\rm s}$ band, though we
did not detect 3-2 S(4) and S(2) in any region. Also, in molecular shock
regions like IC~443 A, B, C, and G here, the $H$ band images can also be
dominant by \ce{H2} ro-vibrational line emission.

\begin{figure*}
\includegraphics[width=1.5\columnwidth]{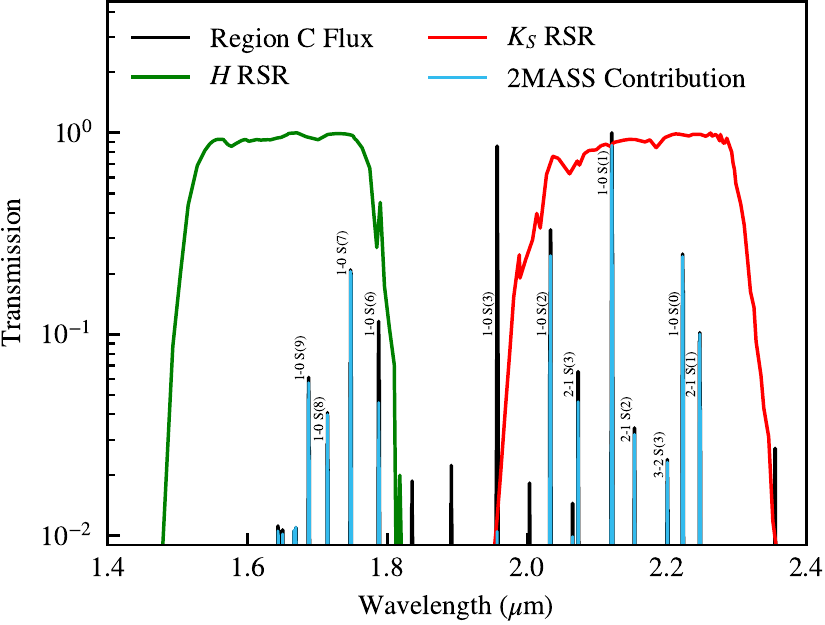}
\caption{Green and Red line: Relative spectral response (RSR) curves of the 2MASS $H$ and $K_{\rm s}$ bands;
Black line: normalized flux of the spectra averaged in region C; Light blue
line: emission line contribution to the 2MASS $H$ and $K_{\rm s}$ band.}
\label{fig:2MASS} \end{figure*}

To get a representative view, we extract a spatially-averaged spectrum from
region C, where the \ce{H2} emission almost distributes all over the FoV
(Figure~\ref{fig:H2Emission}), and normalize to the flux of 1-0 S(1) line
(Figure~\ref{fig:2MASS} black line).  Then we convolve it with 2MASS $H$ and
$K_{\rm s}$ relative spectral response (RSR) \citep[][Figure~\ref{fig:2MASS} green and red
lines]{2003AJ....126.1090C}. The output spectrum is plotted in the light blue line in Figure~\ref{fig:2MASS}. All lines with contribution larger than 1\% of 1-0
S(1) are labeled in the figure.

In the $H$ band, the main \ce{H2} components are 1-0 (from S(6) to S(9))
transitions, with 1-0 S(7) being the strongest one. In the $K_{\rm s}$ band, the
\ce{H2} 1-0 S(1) transition is the strongest one, while the 1-0 S(2), 1-0 S(0),
and 2-1 S(1) transitions contribute at similar levels. The 2-1 S(3) and S(2),
3-2 S(3) transitions also contribute a non-negligible part in the $K_{\rm s}$ band.
Although the 1-0 S(3) transition is almost as strong as 1-0 S(1), it seems to
be attenuated by the low bandpass at the edge.

\subsection{Young stellar object candidates in region G} \label{sec:YSO}

With spectroscopic information from VLT/KMOS, we could identify the young
stellar object (YSO) candidates selected by \cite{2020A&A...644A..64D}. Six of
the 65 YSO candidates selected from the ALLWISE catalog and one of their 79 YSO
candidates selected from the 2MASS catalog are in the coverage of the G Region.
These seven candidates are marked in Figure~\ref{fig:G_para} {\it left} panel
with white (WISE) and grey (2MASS) circles.

All these seven YSO candidates were classified as Class-I YSO. From the
spectral indices derived by \cite{2020A&A...644A..64D}, we can infer these
sources should have a significant flux in $K_{\rm s}$ band, on the order of
$10^{-16}$\,erg\,s$^{-1}$\,cm$^{-2}$\,\AA$^{-1}$, which could be detected with
our sensitivity. 

\begin{figure*}
	\includegraphics[width=2.1\columnwidth]{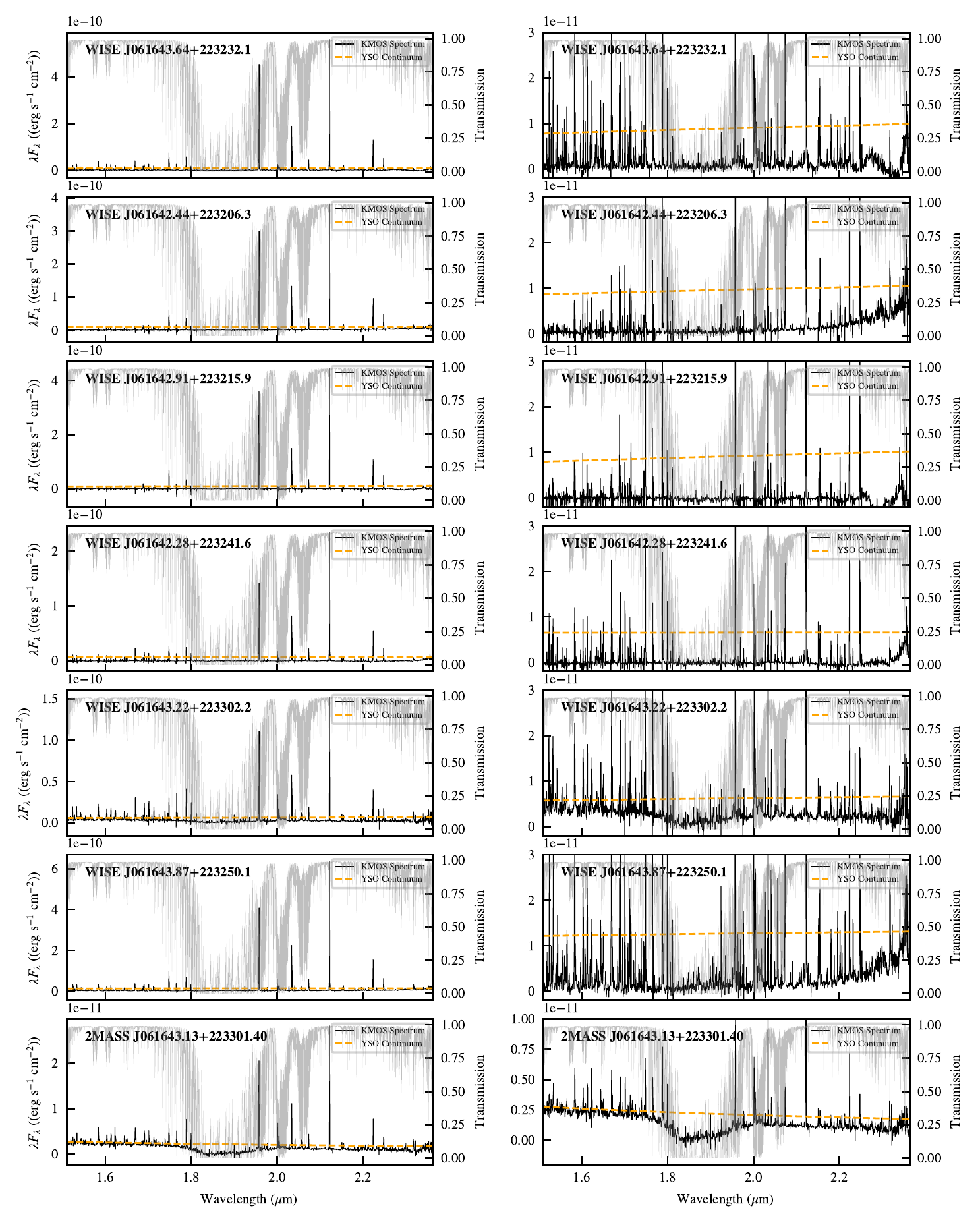}
    \caption{KMOS spectra of the YSO candidates. Orange dashed lines: extrapolated continuum of the YSO candidates; black solid lines: spectra extracted from our data extracted from $r<2{\rm FWHM}$ apertures and subtracted with $2{\rm FWHM}<r<3{\rm FWHM}$ background (The increasing continuum begin from 2.2 $\mu$m is the residuals of molecular emission from the lower atmosphere, refer to Figure~\ref{fig:radiance}); the background grey lines are the transmission at Paranal with PWV=3.5. The \textit{right} column zooms the z-axis of the \textit{left} column.}
    \label{fig:YSO Spectrum}
\end{figure*}

To mock the aperture photometry conducted by WISE/2MASS, we extract the spectra
of these targets by apertures with a diameter of two times the FWHM, and then we
use rings with $2{\rm FWHM}<r<3{\rm FWHM}$ to evaluate and subtract the
background. The FWHMs are $6.4''$ for WISE W2 and $2.5''$ for 2MASS $K_{\rm s}$
images, respectively. In Figure~\ref{fig:YSO Spectrum}, the black lines show
the extracted spectra and the orange dashed lines show the continuum of these
YSO candidates extrapolated from their WISE or 2MASS photometric SED and
spectral index.  The increasing continuum beginning from 2.2 $\mu$m is the
residuals of molecular emission from the lower atmosphere.

%We extract the spectra of these targets by Gaussian tapers with diameter 6.4$''$ and 2.5$''$, which are consistent with the FWHM of WISE W2 and 2MASS $K_{\rm s}$ images, respectively. 

Among them, 2MASS J061643.13+223301.4 overlaps with a star which has a
continuum spectrum. The last row of panels shows that the predicted continuum
is consistent with the observed value. WISE J061643.22+223302.2 is only 1.5$''$
away from this star, so a similar continuum spectrum can also be found in the
fifth-row panels. 

J061642.28+223241.6 is unfortunately located at the edge of our broken IFU,
with one-fourth of the aperture without data.  The other three-fourth aperture
only has \ce{H2} ro-vibrational lines without any continuum contribution.  

All other YSO candidates do not show any continuum in the spectra. Even though
we conducted ring-area background subtraction, we still found strong
nebular emission lines in these spectra. J061642.91+223215.9 and
J061643.87+223250.1 are located in two bright \ce{H2} clumps.
J061642.44+223206.3 and J061643.64+223232.1 also show strong \ce{H2} emission,
and they are only 3.9$''$ and 2.1$''$ away from two of the brightest \ce{H2}
clumps in this FOV, respectively. 

%Therefore, bright \ce{H2} emission dominate the images at $K_{\rm s}$ band.  

We further cross-match the star overlapped with 2MASS J061643.12+223301.4 in
Gaia eDR3 \citep{2021A&A...649A...1G} and LAMOST DR7
\citep{2015RAA....15.1095L} catalogues to check whether it is a young star. We
find one star in a cone of radius 1$''$ and no other stars are found within a
cone of radius 5$''$.

This star has a G magnitude of 16.663 mag, BP-G of 0.726, and G-RP of 0.830.
The parallax in the Gaia catalogue is $0.661\pm0.071$\,mas, and the geometric distance is 
$1451^{+140}_{-150}$\,pc \citep{2021AJ....161..147B}, similar to the distance of IC~443
($\sim1600$\,pc). LAMOST observed this star twice on 21 Nov. 2014 and 21 Jan. 2015, respectively. The spectral type and star parameters including surface
gravity $\log{g}$, and effect temperature $T_{\text{eff}}$ obtained in these
two shots are consistent with each other, while the metallicity [Fe/H] has
a large difference. These observed stellar parameters are
summarised in Table~\ref{tab:starparams}.

If there is any physical connection between this star and SNR
IC~443, this star should be young. 
We use the Bayesian stellar parameter estimator \verb"PARAM 1.5" \footnote{\url{http://stev.oapd.inaf.it/cgi-bin/param}}
\citep{2006A&A...458..609D} and adopt \verb"PARSEC" \citep{2012MNRAS.427..127B}
stellar models to estimate the age of this star with Gaia and LAMOST data. The
most probable ages are $6.60_{-2.14}^{+4.70}$\,Gyr and
$6.02_{-4.17}^{+4.07}$\,Gyr for the LAMOST 2014 and 2015 data, respectively.
This is much older than the progenitor of IC~443, thus it is unlikely to be connected with IC~443 and its progenitor.

\begin{table*}
    \centering
    \caption{Gaia and LAMOST Parameters of 2MASS J061643.12+223301.4.}
    \label{tab:starparams}
    \begin{tabular}{lllllllllll}
   \hline 
        & g Mag.                 & BP-G                   & G-RP                   & Parallax (mas)                   & Distance (pc)                         & Class & $\log{g}$       & {[}Fe/H{]}       & $T_{\text{eff}}$         & Age (Gyr) \\
    \hline 
   2014 & \multirow{2}{*}{16.63} & \multirow{2}{*}{0.726} & \multirow{2}{*}{0.830} & \multirow{2}{*}{$0.661\pm0.071$} & \multirow{2}{*}{$1451^{+140}_{-150}$} & G9            & $4.598\pm0.108$ & $-0.043\pm0.071$ & $4839.52\pm65.82$ & $6.60_{-2.14}^{+4.70}$      \\
   2015 &                        &                        &                        &                                  &                                       & K3            & $4.672\pm0.095$ & $-0.145\pm0.062$ & $4874.03\pm56.28$ & $6.02_{-4.17}^{+4.07}$    \\
    \hline 
        \end{tabular}
\end{table*}

In conclusion, in our observed fields, the majority of the YSO candidates selected from the
2MASS and WISE samples are not continuum sources and their broadband NIR and MIR
images are dominated by the \ce{H2} line emission. The only continuum source we observed is likely an old star whose IR spectrum is significantly contaminated by \ce{H2} line emission as well.

%we conclude that searching YSO candidates in such strong molecular
%shock regions is a sort of tricky because many small bright \ce{H2} clumps or
%knots appear in NIR and MIR bands and look like point sources if observe with
%low resolution imager.

\section{Conclusions} \label{sec:conclusion}

With VLT/KMOS, we observed IC~443 A, B, C, and G, four regions with strong
interaction between supernova shock waves and molecular clumps. Facilitated by
the 8.2-m diameter mirror and near-infrared integral field spectroscopic
instrument KMOS, our observations present twenty \ce{H2} ro-vibrational lines,
Br$\gamma$ line, and two [Fe {\sc ii}] lines in these regions. 

The distribution of warm molecular hydrogen shows clumpy and hierarchical
\ce{H2} structures, with small bright knots that are embedded in the gas
clumps. The molecular hydrogen emission is optically thin at the near-infrared
band and is mainly excited by collisional processes. 

We present the temperature and column density distribution maps and phase
diagrams of these four regions derived by the population diagram method, find
that the temperature and column density of these shocked warm \ce{H2} ranges
from $1000$ K to $4000$ K and from $10^{17}$ cm$^{-2}$ to $10^{19}$ cm$^{-2}$.
These maps give us an intuitive knowledge about how warm molecular gas gets
excited in shocked regions. They can also help us plan and analyze future
multi-band observations zooming in these molecular clumps. 

The extended line profile of [Fe II] 1.6440 $\mu$m line is detected in region A,
which supports that [Fe II] is more likely to come from the fast shock region
where the grain can be sputtered. This also suggests that region A might be in
the early phase of shock-molecular clump interaction when the shock waves
haven't been significantly decelerated yet. 

We detect Br$\gamma$ line in region A with an intensity of
$1.0\pm{0.2}\times10^{-17}$\,erg\,s$^{-1}$\,cm$^{-2}$\,arcsec$^{-2}$ and give a
the new upper limit of Br$\gamma$ line emission in regions B, C, and G of
$1.0\times10^{-17}$\,erg\,s$^{-1}$\,cm$^{-2}$\,arcsec$^{-2}$. Comparing with
the H$\alpha$ emission detected in these regions, we address questions about
the states of the atomic hydrogen in these shocked molecular clumps.

We compare our data with optical, mid-infrared, and radio observations.
The spatial distribution of different phases of gas reveals the shock
structures. Comparing with the total column density of shocked \ce{H2}, we find
that shocked warm \ce{H2} contributes to $<$1\% mass of all the shocked \ce{H2}.

Last, we verified YSO candidates in our Region G data and find no young stellar
counterparts for all of them. The infrared emission of these YSO candidates is
significantly contaminated by \ce{H2} line emission. Cross-checking with
spectroscopic observations thus becomes extremely important for such a YSO
candidate census. 

In the future, multi-band observations of various species in the multi-phase
gas will be wrapped up and give a panchromatic perspective of shocked regions
in IC 443. We have performed Arecibo, JVLA, and FAST observations with a
spatial resolution higher than before. We have also proposed optical integral
field observations with CFHT/SITTILE and large field optical lines narrowband
image of IC 443, and the data will come soon. We also plan to propose high
resolution ($R>50000$) infrared spectroscopic observations with
state-of-the-art telescopes and instruments like VLT/CRIRES+ or CFHT/SPIRou.
These data will help us obtain a more comprehensive understanding of the
stratification of ionization, dissociation structures, and kinematics of
ionized gas, neutral gas, and hot molecular gas in shocked regions.

\section*{Acknowledgements}

This work is based on observations collected at the European Southern
Observatory under ESO programme 0104.C-0924(A). We thank Dr. Paola Popesso and
colleagues from ESO Operation Helpdesk for their kind help and advice. YWD is
grateful to Yichen Sun and Haochang Jiang for their useful discussions.

ZYZ and YWD acknowledge the support of the National Natural Science Foundation of
China (NSFC) under grants No. 12041305, 12173016.  ZYZ and YWD acknowledge the
Program for Innovative Talents, Entrepreneur in Jiangsu.  ZYZ and YWD
acknowledge the science research grants from the China Manned Space Project
with NO.CMS-CSST-2021-A08, NO.CMS-CSST-2021-A07. HL was supported by NASA
through the NASA Hubble Fellowship grant HST-HF2-51438.001-A awarded by the
Space Telescope Science Institute, which is operated by the Association of
Universities for Research in Astronomy, Incorporated, under NASA contract
NAS5-26555. 

We use \textsc{python} packages  \verb"NumPy" \citep{harris2020array},
\verb"SciPy" \citep{2020SciPy-NMeth}, and \verb"Astropy"
\citep{2013A&A...558A..33A,2018AJ....156..123A} to analyze the data cubes and
use \verb"Matplotlib" \citep{Hunter:2007} to visualisation. To remove the
stellar continuum, we use the widely used \verb"Astropy" package
\verb"photutils" to find and mask the stars.

%\ywr{VLT, JVLA, JCMT, IPAC/ZTF, WISE, LAMOST, Gaia}

%%%%%%%%%%%%%%%%%%%%%%%%%%%%%%%%%%%%%%%%%%%%%%%%%%
\section*{Data Availability}

The processed KMOS cubes can be accessed from ESO Science Portal\footnote{\url{http://archive.eso.org/scienceportal/home}}. The raw data and calibration frames of the KMOS observations are provided via the ESO archive facility\footnote{\url{http://archive.eso.org/eso/eso_archive_main.html}}.



\bibliographystyle{mnras}
\bibliography{IC443KMOS} % if your bibtex file is called example.bib

%\begin{thebibliography}{99}
%\end{thebibliography}

%%%%%%%%%%%%%%%%%%%%%%%%%%%%%%%%%%%%%%%%%%%%%%%%%%

%%%%%%%%%%%%%%%%% APPENDICES %%%%%%%%%%%%%%%%%%%%%

\appendix
\section{Atmospheric radiance in KMOS {\it HK} band}
\begin{figure*}
	\includegraphics[width=2\columnwidth]{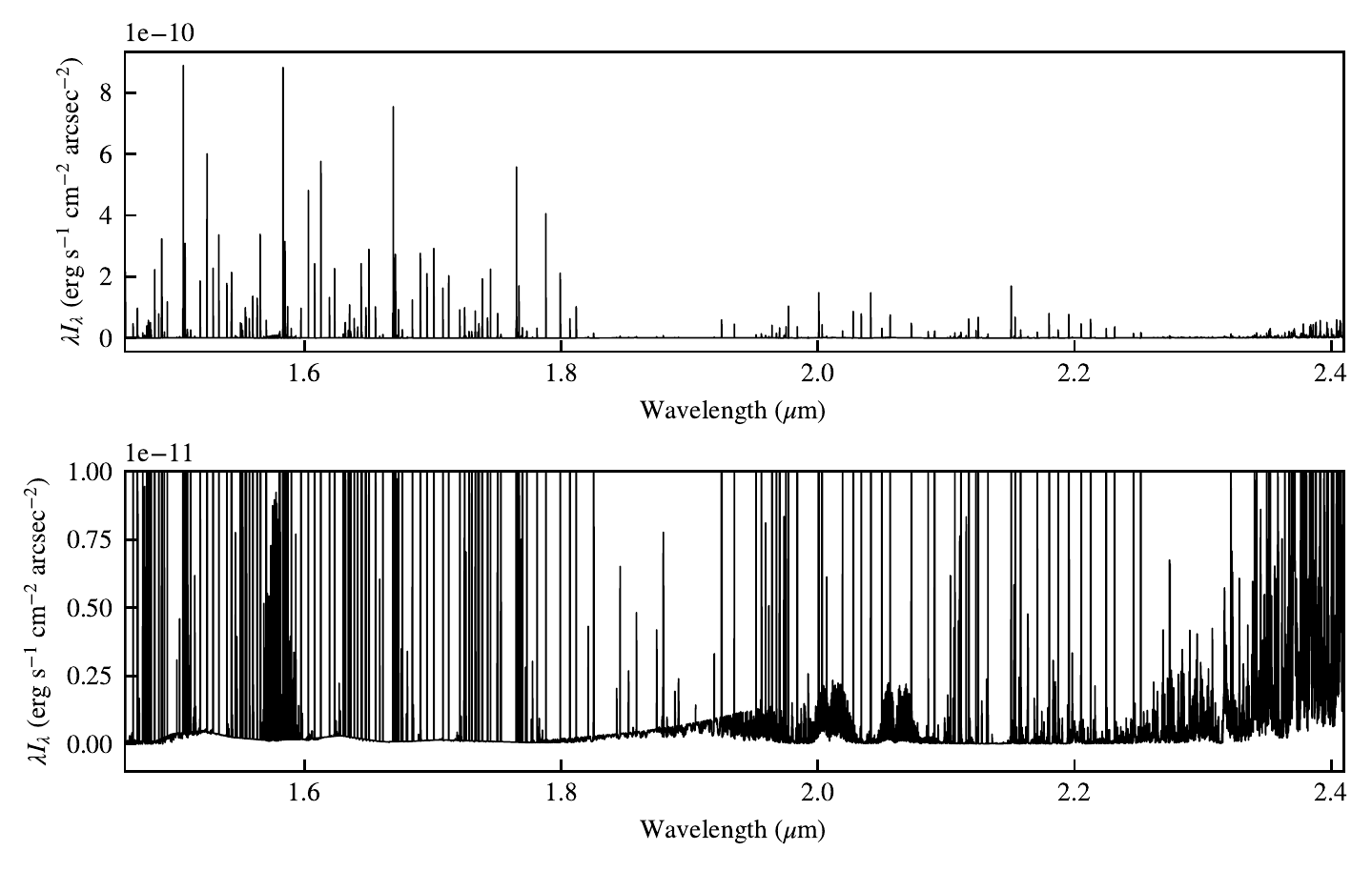}
    \caption{\textit{Top} panel: atmospheric radiance in KMOS {\it HK} band at Paranal with PWV=3.5. The \textit{bottom} panel zooms the z-axis of the \textit{left} column.}
    \label{fig:radiance}
\end{figure*}

The radiance residuals (mainly OH sky lines) are difficult to perfectly subtracted from the spectra. Figure~\ref{fig:radiance} presents the atmospheric radiance spectrum in KMOS {\it HK} band at Paranal with PWV=3.5 obtained from the ESO SkyCalc as a reference.
\section{Small Bright Knots}
Here we list the detected \ce{H2} lines (Table~\ref{tab:H2lines}) in the ten small bright knots.

% \begin{table}
%     \centering
%     \caption{Positions of the 10 small bright knots.}
%     \label{tab:positions}
%     \begin{tabular}{lll}
%     \hline
%         & R.A. (ICRS) & Dec. (ICRS) \\
%     \hline
%     Ac1 & 06:16:54.00  & 22:23:02.30 \\
%     Ac2 & 06:16:54.69  & 22:22:55.55 \\
%     Bc1 & 06:17:16.60  & 22:25:26.00 \\
%     Bc2 & 06:17:16.05  & 22:25:18.25 \\
%     Cc1 & 06:17:42.41  & 22:21:22.35 \\
%     Cc2 & 06:17:42.37  & 22:21:15.15 \\
%     Gc1 & 06:16:43.93  & 22:32:52.65 \\
%     Gc2 & 06:16:43.68  & 22:32:29.65 \\
%     Gc3 & 06:16:42.37  & 22:32:10.05 \\
%     Gc4 & 06:16:43.39  & 22:32:06.65\\
%     \hline
%     \end{tabular}
%     \end{table}

\begin{table*}
    \centering
    \caption{Detected \ce{H2} lines from knots in IC 443 A, B, C, and G (uncorrected for extinction).}
    \label{tab:H2lines}
      \begin{tabular}{lrrrrrrrrrrrr}
        \hline
        \multicolumn{1}{c}{\multirow{2}[0]{*}{Transition}} & \multicolumn{1}{c}{\multirow{2}[0]{*}{$\lambda$ ($\mu$m)}} & \multicolumn{1}{c}{\multirow{2}[0]{*}{$E_{\rm u}/k_{\rm B}$ (K)}} &  \multicolumn{10}{c}{Line Intensity ($10^{-15}$\,erg\,s$^{-1}$\,cm$^{-2}$\,sec$^{-2}$)}\\
        & & &
        \multicolumn{1}{l}{Ac1} & \multicolumn{1}{l}{Ac2} & \multicolumn{1}{l}{Bc1} & \multicolumn{1}{l}{Bc2} & \multicolumn{1}{l}{Cc1} & \multicolumn{1}{l}{Cc2} & \multicolumn{1}{l}{Gc1} & \multicolumn{1}{l}{Gc2} & \multicolumn{1}{l}{Gc3} & \multicolumn{1}{l}{Gc4} \\
              \hline
        1-0 S(11) & 1.6504 & 18979.1 & ...   & ...   & ...   & ...   & 0.10  & 0.26  & ...   & ...   & ...   & ...   \\
        1-0 S(10) & 1.6665 & 17310.8 & 0.12  & 0.10  & ...   & ...   & 0.14  & 0.29  & ...   & ...   & ...   & 0.37  \\
        1-0 S(9)  & 1.6877 & 15721.5 & 0.61  & 0.79  & 0.48  & 0.72  & 1.07  & 1.99  & 1.27  & 0.79  & 0.97  & 0.84  \\
        1-0 S(8)  & 1.7147 & 14220.5 & 0.40  & 0.55  & 0.40  & 0.56  & 0.65  & 1.28  & ...   & 0.72  & 0.90  & 0.57  \\
        1-0 S(7)  & 1.7480 & 12817.3 & 2.10  & 2.90  & 2.35  & 3.46  & 4.15  & 7.31  & 4.40  & 4.50  & 3.90  & 2.78  \\
        1-0 S(6)  & 1.7880 & 11521.1 & 1.12  & 1.53  & 1.41  & 2.16  & 2.19  & 4.06  & 2.71  & 2.83  & 2.03  & 1.54  \\
        1-0 S(5)* & 1.8358 & 10341.2 & 0.13  & 0.32  & 0.30  & 1.13  & 0.11  & 0.37  & ...   & 0.05  & 0.10  & ...   \\
        1-0 S(4)* & 1.8919 & 9286.4  & 0.08  & 0.21  & 0.39  & 1.22  & 0.21  & 0.45  & 0.16  & ...   & 0.09  & ...   \\
        1-0 S(3)* & 1.9576 & 8364.9  & 11.86 & 18.61 & 14.37 & 18.74 & 24.11 & 32.88 & 17.82 & 24.67 & 21.12 & 16.64 \\
        1-0 S(2)  & 2.0338 & 7584.3  & 3.23  & 4.54  & 6.55  & 9.47  & 8.43  & 11.45 & 8.23  & 8.43  & 7.40  & 5.22  \\
        1-0 S(1)  & 2.1218 & 6951.3  & 8.85  & 12.39 & 19.71 & 27.94 & 24.32 & 31.40 & 23.65 & 24.54 & 21.05 & 15.50 \\
        1-0 S(0)  & 2.2233 & 6471.4  & 2.10  & 2.89  & 4.91  & 6.73  & 5.80  & 7.06  & 5.80  & 5.81  & 5.01  & 3.64  \\
        2-1 S(5)* & 1.9449 & 15762.7 & ...   & ...   & 0.13  & 0.29  & 0.06  & 0.12  & 0.07  & 0.08  & 0.05  & ...   \\
        2-1 S(4)* & 2.0041 & 14763.5 & ...   & 0.51  & 0.37  & 0.38  & ...   & 0.30  & 1.27  & ...   & 0.57  & ...   \\
        2-1 S(3)  & 2.0735 & 13890.2 & 0.52  & 0.90  & 1.18  & 1.49  & 1.31  & 2.38  & 2.38  & 2.21  & 1.63  & 0.93  \\
        2-1 S(2)  & 2.1542 & 13150.3 & 0.40  & 0.54  & 0.59  & 0.73  & 0.74  & 1.17  & 0.75  & 0.76  & 0.70  & 0.46  \\
        2-1 S(1)  & 2.2477 & 12550.0 & 1.10  & 1.62  & 1.72  & 2.30  & 2.38  & 3.56  & 2.29  & 2.44  & 2.42  & 1.46  \\
        2-1 S(0)  & 2.3556 & 12094.9 & 0.95  & 0.48  & 0.62  & 0.70  & 0.67  & 1.02  & 0.38  & 0.64  & 0.71  & 0.76  \\
        3-2 S(5)  & 2.0656 & 20855.7 & ...   & 0.23  & ...   & ...   & 0.20  & ...   & ...   & 0.17  & 0.27  & 0.10  \\
        3-2 S(3)  & 2.2014 & 19085.8 & 0.25  & 0.43  & 0.35  & 0.30  & 0.57  & 0.94  & 0.43  & 0.42  & 0.57  & 0.39  \\
        \hline
        \multicolumn{10}{l}{$^*$ These lines are attenuated by the atmospheric absorption. Their intensities are likely underestimated.}\\
        \end{tabular}%
\end{table*}

%%%%%%%%%%%%%%%%%%%%%%%%%%%%%%%%%%%%%%%%%%%%%%%%%%

% Don't change these lines
\bsp	% typesetting comment
\label{lastpage}
\end{document}